\documentclass[conference]{IEEEtran}
\ifCLASSINFOpdf
\else
\fi
\usepackage{fancyhdr}

\usepackage[sort,nocompress]{cite}
\usepackage[hyphens]{url}
\usepackage{graphicx}
\usepackage{textcomp}
\usepackage{xcolor}
\usepackage[bookmarks=true,breaklinks=true,letterpaper=true,colorlinks,linkcolor=black,citecolor=blue,urlcolor=black]{hyperref}

\usepackage[normalem]{ulem}

\usepackage{soul}
\usepackage{algorithm}
\usepackage{xspace}
\usepackage[font=footnotesize,labelfont=bf]{caption}
\usepackage{amsmath}
\usepackage{mathtools}
\usepackage{amsfonts}
\usepackage{amssymb}
\usepackage{pifont}
\usepackage{color}
\usepackage{algpseudocode}
\usepackage{multirow}
\usepackage{textgreek}
\usepackage{subfig}
\usepackage{soul}

\soulregister\cite7
\soulregister\ref7
\soulregister\fig7
\soulregister\todo7
\soulregister\sout7
\soulregister\pageref7
\soulregister\recsys7
\soulregister\emph7

\usepackage{algpseudocode}
\algnewcommand{\LineComment}[1]{\State \(\triangleright\) #1}
\usepackage{tikz}
\newcommand\encircle[1]{
  \tikz[baseline=(X.base)]
    \node (X) [draw, shape=circle, inner sep=-2.0pt, fill=black, text=white] {\strut\scriptsize{#1}};
}
\usetikzlibrary{fit,calc}

\colorlet{pink}{red!40}
\colorlet{blue}{cyan!60}

\def\BibTeX{{\rm B\kern-.05em{\sc i\kern-.025em b}\kern-.08em
    T\kern-.1667em\lower.7ex\hbox{E}\kern-.125emX}}

\newcommand{\ignore}[1]{}
\newcommand{\todo}[1]{\textcolor{red}{#1}\xspace}

\newcommand{\old}[1]{}
\newcommand{\fig}[1]{Figure~\ref{#1}}
\newcommand{\sect}[1]{Section~\ref{#1}}
\newcommand{\tab}[1]{Table~\ref{#1}}
\newcommand{\algo}[1]{Algorithm~\ref{#1}}

\newcommand{\recsys}[0]{RecSys\xspace}
\newcommand{\proposed}[0]{ScratchPipe\xspace}

\newcommand{\RD}[0]{{\bf (R)}\xspace}
\newcommand{\WR}[0]{{\bf (W)}\xspace}

\newcommand\blfootnote[1]{%
\begingroup
\renewcommand\thefootnote{}\footnote{#1}%
\addtocounter{footnote}{-1}%
\endgroup
}

\def\BibTeX{{\rm B\kern-.05em{\sc i\kern-.025em b}\kern-.08em
    T\kern-.1667em\lower.7ex\hbox{E}\kern-.125emX}}

\renewcommand{\baselinestretch}{.999}
\title{\LARGE Training Personalized Recommendation Systems from (GPU) Scratch:
Look Forward not Backwards}

\begin{document}

\author{

\IEEEauthorblockN{
Youngeun Kwon\hspace{2em}Minsoo Rhu}
\IEEEauthorblockA{
School of Electrical Engineering\\
KAIST\\
\texttt{\{yekwon, mrhu\}@kaist.ac.kr}\\
}
}

\maketitle
\pagestyle{plain}

\begin{abstract}

Personalized recommendation models (\recsys) are one of the most popular
machine learning workload serviced by hyperscalers. A critical
challenge of training \recsys is its high 
 memory capacity 
requirements, reaching hundreds of GBs to TBs
of model size. In \recsys, the so-called embedding layers
account for the majority of memory usage so current systems
employ a hybrid CPU-GPU design to have the large CPU memory store
the memory hungry embedding layers. Unfortunately, training embeddings
involve several memory bandwidth intensive operations which is at odds
with the slow CPU memory, causing performance overheads. Prior work proposed
to cache frequently accessed embeddings inside GPU memory as means to 
filter down the embedding layer traffic to CPU memory, but this paper observes
several limitations with such cache design.
In this work, we present a fundamentally different approach in designing 
embedding caches for \recsys. Our proposed \proposed architecture utilizes
unique properties of \recsys training to develop an embedding  cache 
that not only sees the past but also the ``future'' cache accesses. \proposed
exploits such property to guarantee
that the active working set of embedding layers can ``always'' be captured
inside our proposed cache design, enabling embedding layer training 
to be conducted at GPU memory speed. 

\end{abstract}

\IEEEpeerreviewmaketitle
\blfootnote{
This is the author preprint version of the work. The authoritative version will appear in the Proceedings of the $49^{\text{th}}$ IEEE/ACM International Symposium on Computer Architecture (ISCA-49), 2022.
}

\section {Introduction}
\label{sect:intro}

Personalized recommendation systems (\recsys) are one of the most successfully
commercialized machine learning (ML) applications in the industry, widely being
deployed for online services (e.g., video recommendations from
		YouTube/Netflix, product recommendations from Amazon).  Recent literature 
observes that \recsys also
follows the ``larger the model, the better the accuracy'' principle, a
study from Facebook claiming their production \recsys have increased by
tenfold within the past three
years~\cite{lui2021understanding,mudigere2021high}.  Such scaled-``up'' model
development trends rendered state-of-the-art \recsys to reach several
hundreds of GBs to TBs of model size~\cite{aibox,
	yi2018factorized,mudigere2021high}.

	Unlike ML systems for computer vision or natural language processing where GPUs have been the preferred architecture of
	choice for training, \recsys model's unique memory requirements make
	CPUs play a critical role in training these
	models. In \recsys, the so-called \emph{embedding tables} account for
		the majority of memory usage, which is a large lookup table that stores
		millions to billions of embeddings~\cite{lui2021understanding}.
High-end GPUs~\cite{ampere_a100} or TPUs~\cite{tpuv4}
		tailored for ML training typically employ high-bandwidth
		but low-capacity memory solutions like HBM~\cite{hbm}, whose
		(only) several tens of GBs of storage falls short in capturing the
		enormous working set of embedding tables. Consequently, system
		architectures for training \recsys typically employ a \emph{hybrid} CPU-GPU
		system.  At a high-level, \recsys models
		consist of two parts, the frontend embedding layers and the backend DNN
		layers.  Under the CPU-GPU system, the CPU stores the embedding
		tables inside capacity-optimized CPU memory, thus handling the training
		process of frontend embedding layers.  The GPU then handles the training
		process of the backend DNN layers using its 
		high-bandwidth (but capacity-limited) memory.

		A {\bf key limitation} of training embedding layers over the CPU memory is that
		the majority of its execution time is spent conducting a highly memory
		bandwidth limited operation, which is at odds with the low-throughput CPU
		memory.  An embedding layer conducts the following key primitives
		during forward propagation (and backpropagation), an embedding
		\emph{gather} operation from the embedding tables (and gradient
				\emph{scatter} operation to the table) followed by a \emph{reduction}
		operation among the gathered embeddings, all of which are highly memory
		intensive. Consequently, training embedding layers over the CPU-GPU system
		is known to incur performance bottlenecks~\cite{tcasting}.

To address these challenges, recent work proposed to leverage \emph{locality}
inherent in the memory accesses to/from embedding
tables~\cite{merci,recnmp,yonsei_space,ttrec}, seeking to alleviate
embedding layer's memory throughput requirement. Prior work observes that
embedding table accesses follow a power-law distribution, i.e., a small subset
of ``hot'' embeddings capture a significant portion of the overall accesses to
the embedding table.  Based on such insight, a CPU-GPU system augmented with a
software-managed \emph{GPU embedding cache} can store the most frequently
accessed hot embeddings in fast GPU memory, which helps reduce the memory
bandwidth bottlenecks of CPU-side embedding layer training.  As we detail in
\sect{sect:motivation_caching}, however,  a CPU-GPU system augmented with a GPU
embedding cache still suffers from non-trivial amount of  embedding cache
misses.  Under such circumstances, the CPU-side embedding tables must
inevitably be accessed to retrieve the \emph{missed} embeddings, the latency of
which sits in the critical path of the end-to-end training, hurting
performance.

In this work, we present a fundamentally different approach in designing
software embedding caches for \recsys training. 
Conventional caches
typically employ a history based, \emph{speculative} cache
insertion/replacement policy. This is because it is generally impossible to know
what memory accesses will be observed in the future, so caches are generally designed to utilize past
cache access patterns to make educated guesses on what \emph{may} happen moving forward.
Under the context of
training \recsys, we make the {\bf key observation} that it is actually
possible to ``precisely'' know when and how many memory accesses to the
embedding tables will occur in the future.  This is because the \emph{training
	dataset} records exactly which indices to utilize for 
	embedding gathers (and gradient scatters) as these will be the primary target
	for model updates during training, not just for the current but also for all
	upcoming training iterations. 
Our proposal
utilizes such insight to identify what embedding table accesses
will occur in future training iterations and utilize that information to
develop an \emph{optimal} GPU embedding cache that ``always hits''.

To this end, we propose \proposed, a software runtime system that manages
high-bandwidth GPU DRAM as a fast ``scratchpad\footnote{Note that \proposed's scratchpad memory is managed in GPU
	``DRAM'', which is completely different from CUDA's \emph{shared memory}
	(which is a CUDA programmer managed	scratchpad space stored in on-chip
	 SRAMs).}''. 
Our software runtime  is carefully designed to systematically
bring in the required embeddings from CPU to our GPU scratchpad
right before the embedding training procedure is initiated such that
embedding table accesses always become hits.
	 \proposed first fetches soon-to-be-accessed embedding
	 table lookup IDs from the training dataset, completely transparent to both
the programmer as well as the ML framework (e.g., PyTorch).  The
fetched lookup IDs are then utilized by the \proposed 
controller to identify the set of embeddings to proactively \emph{prefetch}
from CPU memory and subsequently copied over to GPU scratchpad.
		To \emph{hide} the
 latency overhead of CPU$\rightarrow$GPU embedding prefetching,
\proposed
				collects \emph{multiple} mini-batches  from the training dataset, concurrently processing
				different stages of multiple training iterations via 
``pipelined'' execution.
	Because \proposed guarantees the
	prefetched embeddings are filled into the GPU scratchpad before the
	on-demand cache accesses are initiated, our proposal provides the illusion of
	a ``GPU-only'' system (designed over the baseline CPU-GPU) that
	enables embedding table accesses to occur at GPU memory speed.
As such, \proposed can achieve commensuate performance to a multi-GPU based, 
model parallel	GPU-only system 
using only a single GPU machine, significantly reducing \recsys training cost.

	Overall, \proposed achieves an average $5.1\times$ (max $6.6\times$) and
	$2.8\times$ (max $4.2\times$) speedup vs. the baseline hybrid CPU-GPU without
	and with software-managed GPU embedding caches, respectively.
	Furthermore, while \proposed only utilizes a single GPU machine for training,
	we achieve comparable performance to an $8$ multi-GPU system that can	store
	all the embedding tables entirely within GPU's high-bandwidth memory (i.e.,
			GPU-only), thus providing a significant reduction in per epoch training
	cost (avg $4.0\times$, max $5.7\times$), enabling cost-effective \recsys
	training in cloud datacenters.

\section{Background and Related Work}
\label{sect:background}

\subsection{RecSys Models and Embedding Layers}
\label{sect:background_recsys}

\begin{figure}[t!] \centering
    \includegraphics[width=0.475\textwidth]{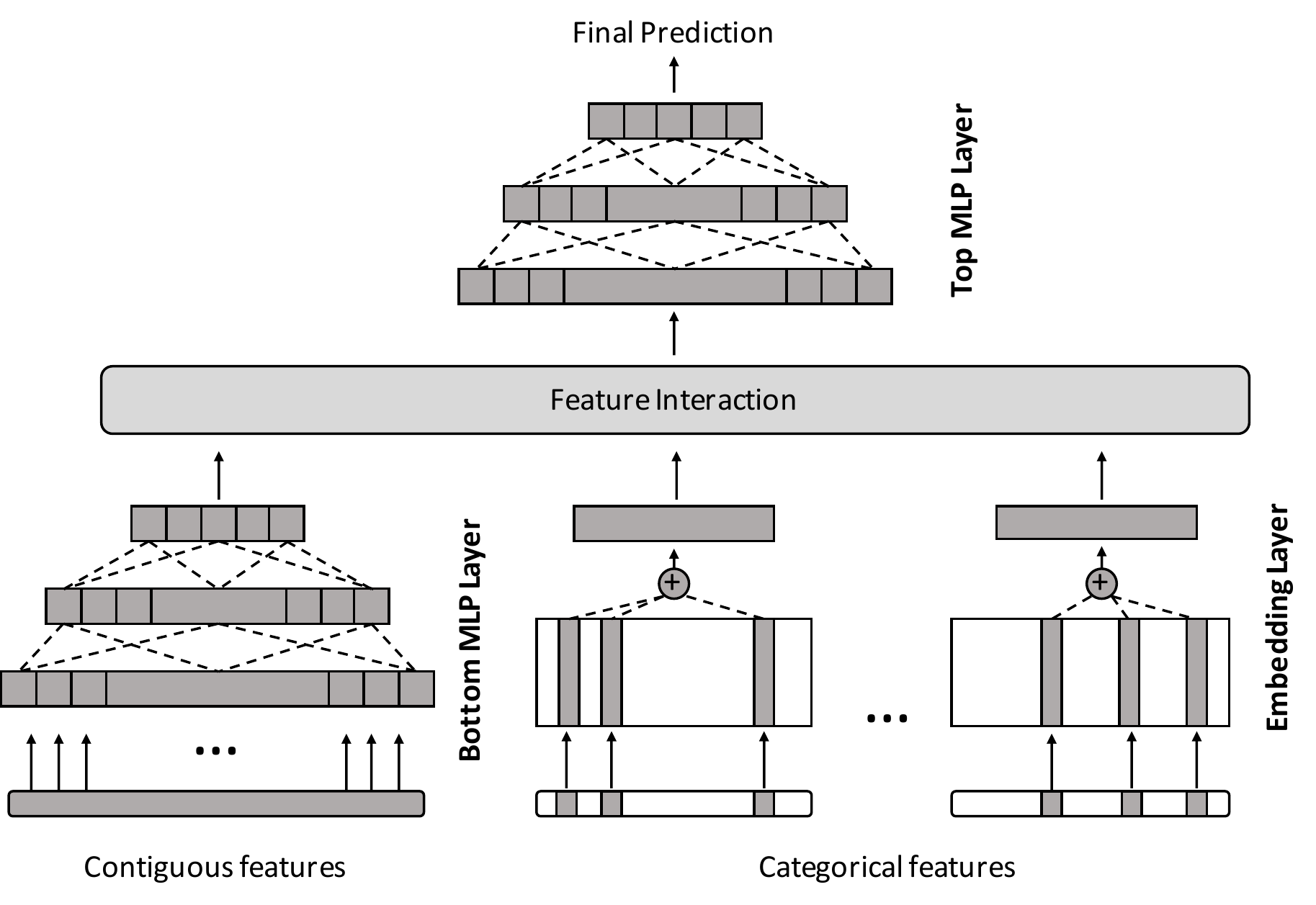}
 \vspace{-0.5em}
\caption{
High-level overview of a neural network based \recsys model.
}
\vspace{-0.5em}
\label{fig:recsys_model}
\end{figure}

\fig{fig:recsys_model} shows a DNN based \recsys, which is composed of two
major components: the \emph{embedding layer} and the DNN layer using
\emph{multi-layer perceptrons} (MLP).  \recsys is typically formulated as a
problem of predicting the probability of a certain event (e.g., the likelihood
		of an e-commerce shopper to purchase the recommended product), so the
accuracy of \recsys depends on how the unique properties of various input
features are captured.  An \emph{embedding} is a projection of a
discrete-valued, categorical feature into a vector of continuous real-valued
numbers. For instance, different product items sold in e-commerce services are
projected from a discrete ID (i.e., each project item is assigned a unique ID
		for differentiation) into a continuous, real-valued vector representation.
Such process is done by \emph{training} the embeddings using embedding layers.
Because the number of unique items falling under a particular feature typically
amounts to several millions to billions (i.e., the number scales proportional
		to the number of products sold in e-commerce or videos available in online
		video streaming services), an \emph{embedding table} which stores all
embeddings amounts to several tens of GBs of capacity.  Because there can be
multiple categorical features (e.g., user ID, item ID, $\ldots$) that are
helpful in capturing semantic representations, several tens of embedding tables
can be utilized per \recsys, rendering the  overall model to consume several
hundreds of GBs to even TBs of memory capacity~\cite{aibox, yi2018factorized}.

\subsection{Training Pipeline for RecSys}
\label{sect:background_training}

\begin{figure}[t!] \centering
	\vspace{-1.3em}
    \subfloat[]
    {
        \includegraphics[width=0.45\textwidth]{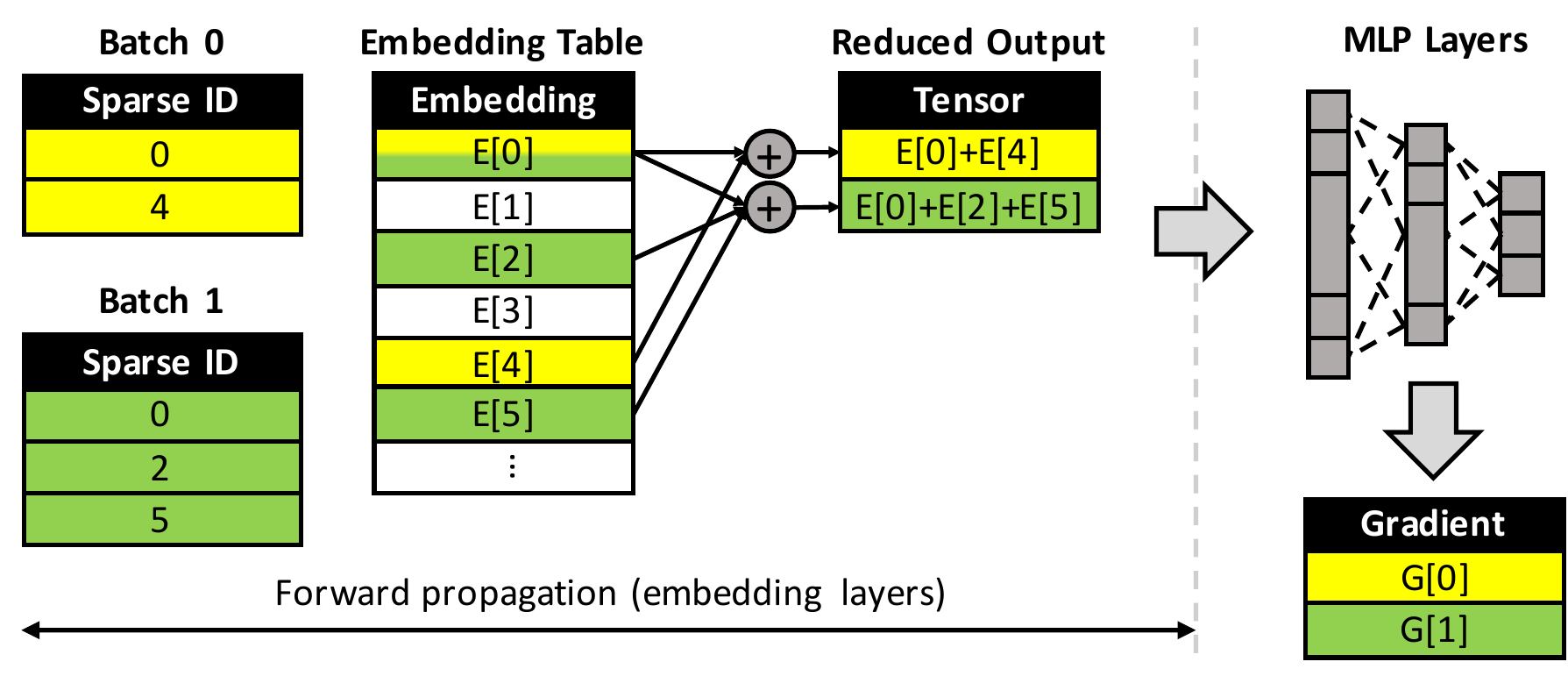}
    }
    \vspace{-0.5em}
    \subfloat[]
    {
    \includegraphics[width=0.45\textwidth]{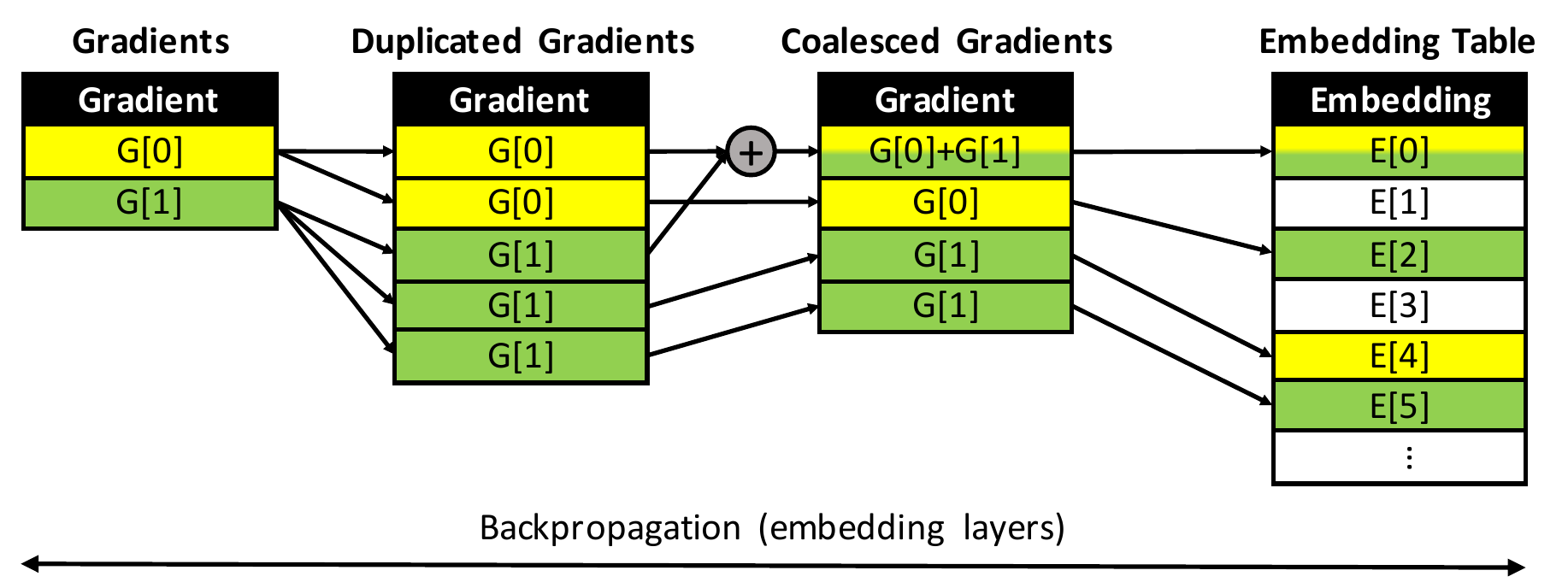}
    }
    \vspace{0em}
    \caption{
    High-level overview of (a) forward and (b) backpropagation of embedding
            layers. Example assumes a batch size of $2$, each batch input gathering $2$
            and $3$ embeddings, respectively. The sparse IDs
							of all batches, used for gather/scatter, are stored
						as part of the \emph{training dataset}.
    }
    \label{fig:recsys_training}
		\vspace{-1.3em}
\end{figure}

{\bf Forward propagation.}
\fig{fig:recsys_training}(a) shows the key computations conducted during 
forward propagation of training the embedding layers in \recsys.
	Using the embedding tables, 
	a discrete, categorical feature is 
	mapped into its corresponding vector representation via embedding 
	\emph{gather} operations, i.e., utilize a group of sparse feature IDs to index the
	embedding table and read out the corresponding embeddings. The group of IDs
	used for gathering embeddings do not necessarily point to contiguous rows within the
	embedding table, so an embedding gather operation exhibits a highly \emph{sparse} and
	\emph{irregular} memory access pattern, exhibiting a \emph{memory bandwidth limited} property. The gathered embeddings
	are combined with each other via element-wise operations, hence \emph{reduced}
	down to a single vector. The embedding reduction is done per
	each table and the reduced embeddings are combined with 
	the output of bottom MLP layers (whose role is to transform continuous input features
	into an intermediate feature vector) via the feature interaction stage. 
	The output of feature interaction is handled by the top MLP layer and subsequently processed using
	softmax functions to predict the final click through rate (CTR), e.g., the probability of
	an end-user to click the recommended item. The backpropagation stage of MLP layers then
	derives a set of \emph{gradient} vectors that are routed back to the embedding layers. Note
	that the number of gradients to backpropagate is identical to the number of reduced
	embeddings during forward propagation (e.g., two in \fig{fig:recsys_training}(a), \texttt{G[0]}, \texttt{G[1]}).

{\bf Backpropagation.} An interesting property of \recsys training is that the
embeddings stored inside embedding tables are the ones subject for model
updates, i.e., embedding tables are both read and written during the course of
training.  Specifically, during a single iteration of forward and
backpropagation, it is those \emph{gathered} embeddings during forward
propagation that will be the target of training during backpropagation.  This
process is illustrated in \fig{fig:recsys_training}(b), where the two
gradients, backpropagated from the backend MLP layers, are used to update
multiple rows within the embedding table (i.e., those looked up locations that
		have been gathered during forward propagation) using a gradient
\emph{scatter} operation. For instance, in \fig{fig:recsys_training}, the
embedding table's rows $0$ and $4$ for the first batch and 	rows $0$, $2$, and
$5$ for the second batch were gathered during forward propagation (a), so these
locations are targeted for gradient scatter update during backpropagation (b).
Because any given embedding can be read out \emph{multiple} times across
\emph{different} batches (e.g., \texttt{E[0]} at row $0$ is looked up twice for
		the first and second batch), multiple backpropagated gradients must be
accounted for when deriving the final gradients to use for model updates. This
is handled by a series of gradient \emph{duplication} and \emph{coalescing},
	 which is also a memory bandwidth limited operation.

\subsection{System Architectures for Training \recsys}
\label{sect:background_sysarch}

State-of-the-art \recsys employ large embedding tables which can require up to
several TBs of memory (\sect{sect:background_recsys}).  Since training is a
throughput-bound operation, high-end GPUs or TPUs employ bandwidth-optimized
but capacity-limited memory solutions like 3D stacked DRAMs (e.g.,
		HBM~\cite{hbm}). As these memory solutions only come with several tens of
GBs of storage, it becomes very challenging to store the enormous embedding
tables within GPU local memory. Consequently, system architectures for \recsys
training typically adopt a hybrid CPU-GPU system (e.g., Facebook's Zion
		system~\cite{zion}, Baidu's AIBox~\cite{aibox}). In such system design
point, the capacity-optimized CPU DIMMs are used to store the embedding tables
so that the CPU handles the training of memory-intensive embedding layers,
	 whereas the GPU conducts the training of compute-intensive DNN layers.
	 Since the key compute primitives of both forward and backpropagation of
	 embedding layers are highly memory bandwidth limited
	 (\sect{sect:background_training}), executing them over the slow CPU memory
	 causes significant slowdown in \recsys training.

\begin{figure}[t!] \centering
    \includegraphics[width=0.48\textwidth]{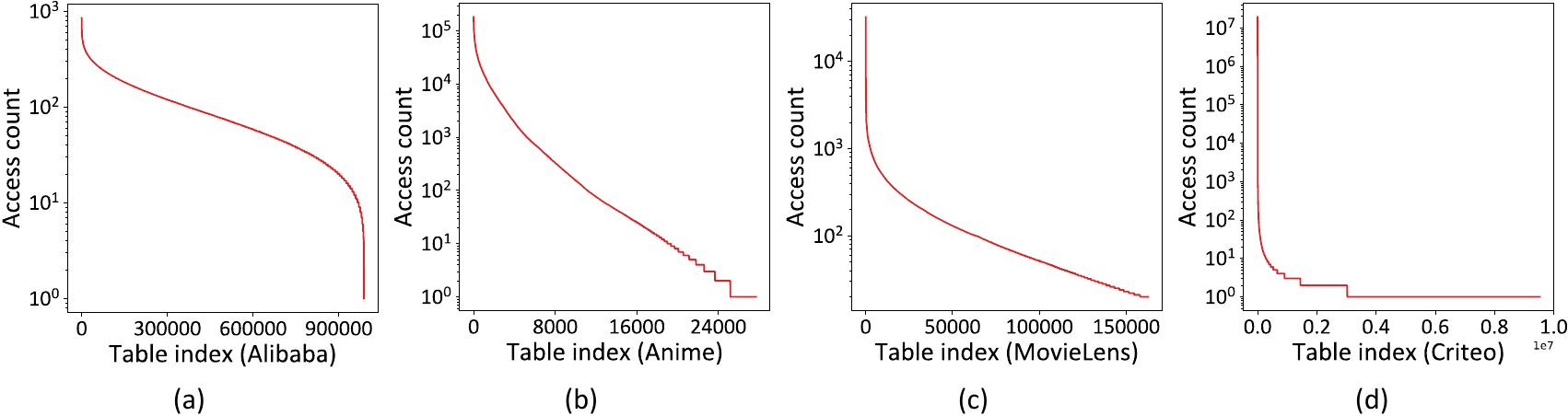}
\caption{
A (sorted) access count of embedding table entries in \recsys datasets: (a) Alibaba, (b) Kaggle Anime, (c) MovieLens, and (d) Criteo.
}
\vspace{-.5em}
\label{fig:motivation_locality}
\end{figure}

\begin{figure*}[t!] \centering
    \includegraphics[width=0.995\textwidth]{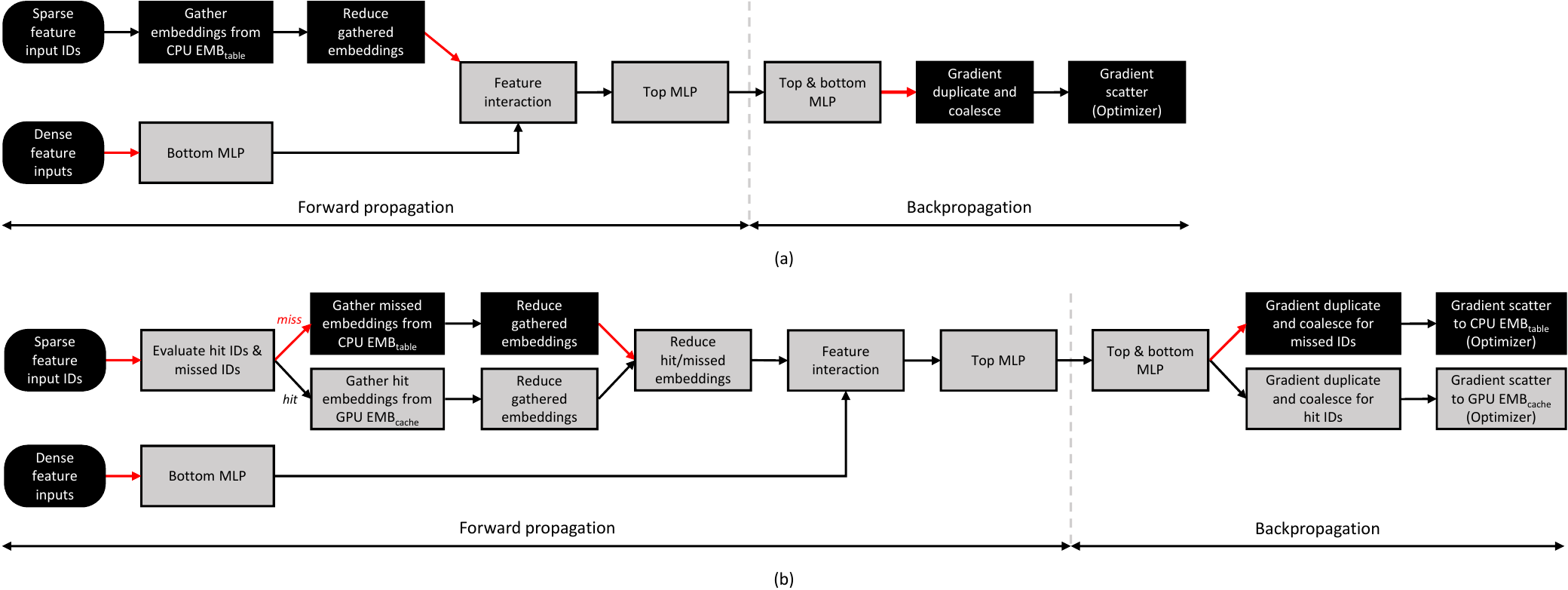}
\caption{
\recsys training pipeline under (a) baseline CPU-GPU without caching~\cite{tcasting}, and
	(b) CPU-GPU augmented with a software-managed GPU embedding cache (EMB$_{cache}$) that statically
	stores top-$N$
	most-frequently-accessed embeddings of the original embedding table (EMB$_{table}$) stored in CPU memory~\cite{ttrec}. In (b), the sparse feature IDs are first copied over to the GPU to evaluate hit/missed IDs. The missed IDs are then copied back to CPU memory for embedding table lookups. Later during the backpropagation of embedding layers, the embeddings corresponding to hit IDs (missed IDs) are updated in EMB$_{cache}$ over GPU memory (in EMB$_{table}$ over CPU memory). Gray (black) stages are those executed on the GPU (CPU). Red arrows designate CPU$\leftrightarrow$GPU copies.
}
\vspace{-.5em}
\label{fig:baseline_vs_caching_flow}
\end{figure*}

\subsection{Related Work}
\label{sect:related}

Unlike prior literature focusing on the acceleration of compute-intensive DNN
algorithms~\cite{tpu1,scnn,cambricon,eyeriss_isca,cnvlutin,diannao,song:2015:eie,maeri,dadiannao,stripes,neurocube},
					 \recsys models pose a unique challenge for computer architects as
					 they incur unprecedented levels of memory capacity and bandwidth
					 demands~\cite{dlrm:arch}.  TensorDIMM, RecNMP, Centaur, Fafnir, and
					 TRiM~\cite{tensordimm,recnmp,centaur:hwang,fafnir,kim2020trim,trim} propose a near-/in-memory processing solution
					 specialized for embedding layer's gather-and-reduce to overcome the
					 memory bandwidth limitations of this primitive under \recsys
					 inference.  More recently, RecSSD explored the efficacy of employing
					 SSDs to cost-effectively store TB-scale embedding
					 tables~\cite{recssd}. To close the wide performance gap between DRAM
					 and SSD, RecSSD employs an in-storage processing unit for embedding
					 gather-and-reduce.  Under the context of training, Tensor
					 Casting~\cite{tcasting} proposed a near-memory processing
					 architecture for \recsys's embedding layer training.

Because near-/in-memory or in-storage processing requires modification to the
current hardware/software stack, recent studies explored the possibility of
leveraging locality inherent in the read/write accesses from/to embedding
tables to cost-effectively reduce embedding layer's memory bandwidth
demands~\cite{recnmp,recssd,ttrec,yang2020mixed,merci,yonsei_space}.  RecNMP
and RecSSD both observe that a small subset of embedding table entries account
for a significant fraction of embedding table lookups, exhibiting a power-law
distribution.  Therefore, these studies employ a lightweight
near-memory/in-storage \emph{embedding cache} that helps filter out  the
embedding read traffic for the ``hot'' embedding table entries.
MERCI~\cite{merci} and SPACE~\cite{yonsei_space} similarly observe locality
within the gather and reduction operations of \recsys, proposing memorization~\cite{merci}
and a heterogeneous memory hierarchy~\cite{yonsei_space} respectively for fast
\emph{inference}.  There is also recent work that co-designs the \recsys
training algorithm with the underlying hardware/software system, reducing
memory capacity~\cite{ttrec} or memory bandwidth demands~\cite{yang2020mixed}
of \recsys training. For instance, Yin et al.~\cite{ttrec} suggests matrix
decomposition and approximation techniques to reduce the memory size of
embedding layers.  As these techniques add more computation and latency
overheads, Yin et al.  proposed a software-managed, GPU embedding cache that
statically caches the top-$N$ hot entries as means to filter down the number of
high overhead matrix multiplications.  Yang et al.~\cite{yang2020mixed} explore
a mixed precision training system that seeks to balance \recsys's memory
capacity requirements and training speed.  Our \proposed does not change the
algorithmic properties of \recsys training and provides \emph{identical}
training accuracy vs. the original training algorithm executed over baseline
hybrid CPU-GPU. There is also a large body of prior work that utilizes
	software-level pipelining to overlap computation with data movement as means
		to hide CPU-GPU communication
		latency~\cite{marius,aibox,zeroinfinity,saberlda,xiedistributed,scalefreectr,pipedream}.
		The key
		contribution of \proposed and its applicability is orthogonal to 
		these prior literature as they either target a different algorithm, assume a different
		heterogeneous memory system, or
		are driven by different observations. The uniqueness of \proposed lies in its fine-grained,
		multi-stage software pipelined architecture tailored to  the CPU-GPU memory system, enabling
		the concurrent processing of multiple input mini-batches within its pipeline.
		We observe, however, that such parallel processing of multiple mini-batches invoke complex data hazards so
		\proposed employs a novel hazard resolution mechanism to guarantee that the algorithmic nature of \recsys
		training is not altered.
While not specifically focusing on recommendation models, 
there is a rich set of prior literature exploring heterogeneous memory systems for 
training large-scale ML algorithms~\cite{rhu:2016:vdnn,rhu:2018:cdma,mcdla:cal,mcdla,kwon:2019:disagg,
capuchin,ibm:lms:2018,swapadvisor,layer_centric:taco,sentinel,superneurons}.
In general, the key contribution of our \proposed is orthogonal to these prior studies.

\section{Motivation} 
\label{sect:characterization}
\old{
In this section, we conduct a detailed characterization on the
locality in embedding table accesses which the 
baseline hybrid CPU-GPU system can exploit for \recsys training
performance improvements. 
training performance, root-causing its performance bottlenecks.  
as discussing the benefits of utilizing caching for training system performance
improvements.
}

\subsection{Locality in Embedding Layer Training}
\label{sect:motivation_locality}

In \fig{fig:motivation_locality}, we show the (sorted) access count of  
embedding table entries in various real-world \recsys datasets.
The embedding table accesses generally exhibit a power-law
distribution where a small subset of table entries receive very high access
frequency while the remaining entries receive only a small number of accesses.
Such long-tail phenomenon  is not surprising as it is typically the
case where a large fraction of \recsys users are interested in a small portion
of the most popular items (e.g., popular products in Amazon or most watched
		movies in Netflix). While the magnitude of locality  is highly dependent on where the \recsys
		model is being deployed for (e.g., in Criteo Ad Labs, $2\%$ of the
				embeddings account for more than $80\%$ of all accesses whereas
				for Alibaba User dataset, $2\%$ of embeddings ``only'' account
				for $8.5\%$ of traffic), we generally observe that \recsys 
		datasets indeed exhibit the power-law distribution with a very long-tail.

Unfortunately, the baseline hybrid CPU-GPU is not designed to leverage
the unique locality properties of embedding layers~\cite{tcasting}.
\fig{fig:baseline_vs_caching_flow}(a) shows the key stages of forward and
backpropagation under baseline CPU-GPU. As depicted,  the memory bandwidth limited
embedding gather and gradient scatter operations are all conducted over the
slow CPU memory, so the end-to-end training time is primarily
bottlenecked on the CPU-side training of embedding layers, consuming
significant portion of training time
(\fig{fig:motivation_latency_breakdown}).

\subsection{Software-Managed Embedding Caches}
\label{sect:motivation_caching}

			\begin{figure}[t!] \centering
    \includegraphics[width=0.48\textwidth]{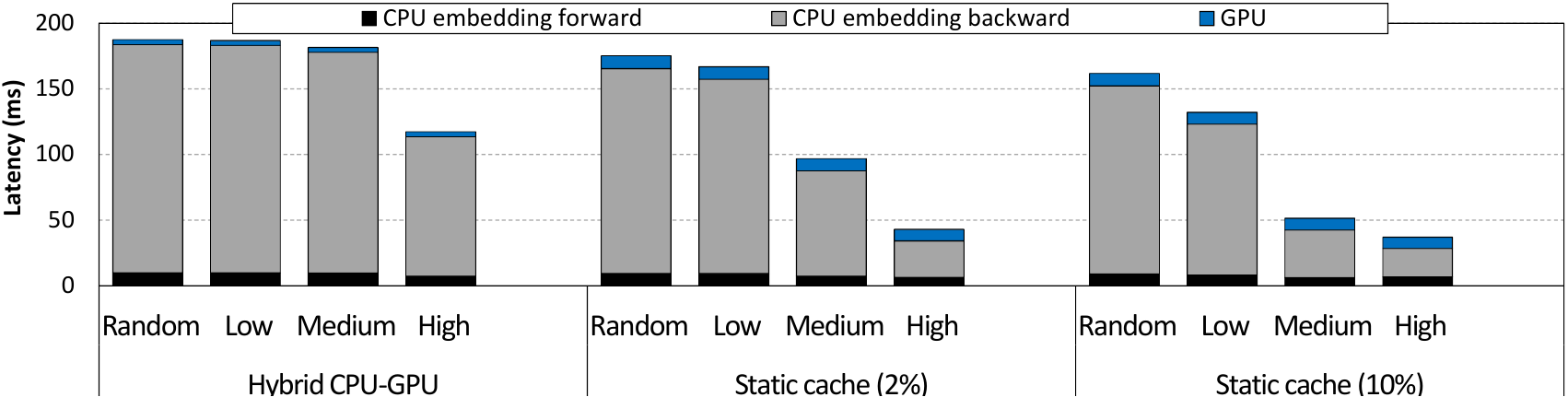}
    \caption{
      Training time broken down based on where key stages are executed (CPU vs. GPU).
				The software-managed GPU embedding cache is sized to store the
				most-frequently-accessed top $2\%$/$10\%$ of embedding table entries.
				Random/Low/Medium/High in the x-axis indicates the magnitude of locality
				inherent in the \recsys dataset we study.
				\sect{sect:methodology} details our methodology.
        }
        \label{fig:motivation_latency_breakdown}
\end{figure}

Given the power-law distribution of embedding table accesses, an effective
mechanism to exploit such locality is to incorporate a small GPU
\emph{embedding cache} (managed in fast GPU DRAM) that can help reduce the
embedding gather/gradient scatter traffic to CPU memory.
\fig{fig:baseline_vs_caching_flow}(b) provides an overview of \recsys training
under a CPU-GPU system enhanced with a software-managed GPU embedding cache.
The key data structures of such software-level cache (e.g., tag, data, other metadata)
	as well as its cache controller module
are all implemented and managed in software
	over GPU memory using CUDA~\cite{cuda}. In 
	terms of cache management policies, we
assume the \emph{static} cache architecture suggested by Yin et
al.~\cite{ttrec} where the most-frequently-accessed embeddings (i.e., the
top-$N$ leftmost table entries in \fig{fig:motivation_locality}) are chosen
for caching in GPU DRAM without getting evicted throughout the entire \recsys
training process. 
\fig{fig:motivation_latency_breakdown} shows the normalized
end-to-end training time broken down based on where the key stages of training
is executed.  We make several important observations from this experiment.
First, the static GPU embedding cache noticeably reduces the time spent in the
memory bandwidth limited embedding layer training. This is because of the hot
embeddings hitting in the GPU embedding cache, which helps filter out the
number of embedding table accesses serviced by the CPU memory.  Unfortunately,
			 there is still a non-negligible amount of time spent in servicing the
			 (embedding cache missed) embedding gathers and gradient scatter
			 operations over CPU memory, averaging at $77\%$ to $94\%$ of
training time. In particular, the backpropagation of embedding layers
	corresponding to the \emph{missed IDs} (the rightmost two black stages in
			\fig{fig:baseline_vs_caching_flow}(b))	invoke memory bandwidth
		limited gradient duplication/coalescing/scatters over the slow CPU
		memory, causing serious latency overheads. 
	 On
		average, the static GPU embedding cache suffers from $12\%$ (high locality dataset) to
			$91\%$ (low locality dataset) cache miss rates, all of which must be serviced from CPU
			embedding tables.
		
	\begin{figure}[t!] \centering
        \includegraphics[width=0.48\textwidth]{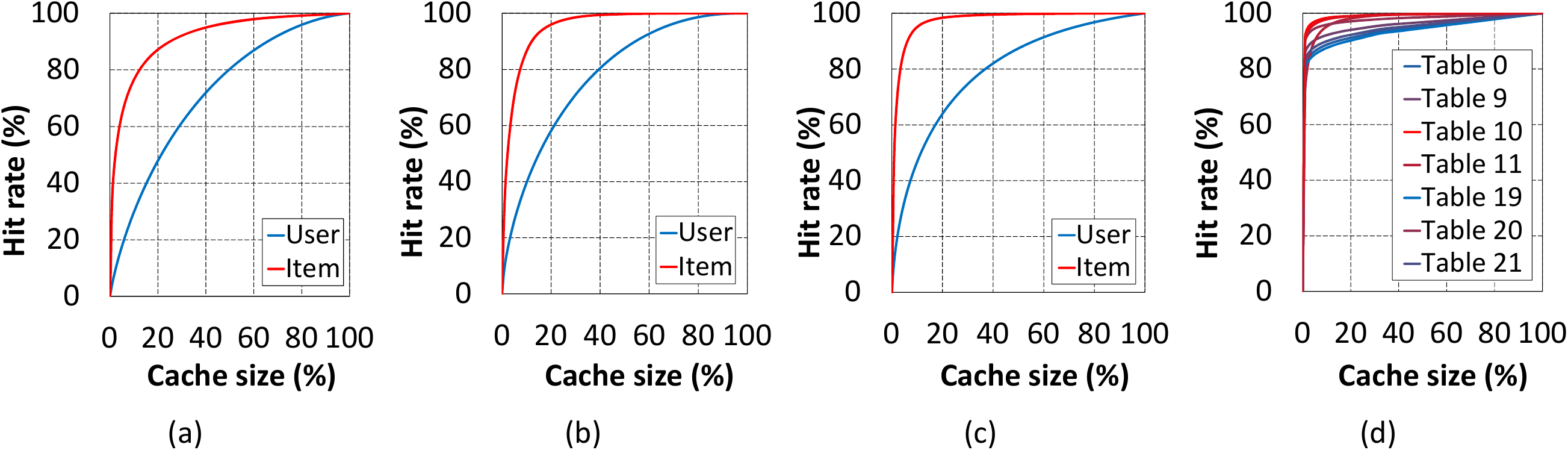}
    \caption{
			Static GPU embedding cache hit rate as a function of cache size, for
				(a) Alibaba, (b) Kaggle Anime, (c) MovieLens, and (d) Criteo. $100\%$ cache size
				means the GPU embedding cache is large enough to fully cache the entire embedding
				tables within GPU's local high-bandwidth memory. 
        }
        \vspace{-1.25em}
    \label{fig:motivation_cache_hitrate}
\end{figure}

	One might argue that a \emph{larger} GPU embedding cache could
	potentially absorb the majority of traffic to CPU memory and enable the
	performance of CPU-GPU to reach that of a ``GPU-only'' system (i.e., the
			high-bandwidth GPU memory stores \emph{all} embedding tables, allowing
			embedding layers to be trained at  GPU memory speed).
	\fig{fig:motivation_cache_hitrate} plots the improvements in GPU embedding
	cache hit rate as we cache a larger fraction of CPU embedding tables. For
	datasets like Criteo, a small fraction of hot embedding table entries exhibit
	extremely high locality (\fig{fig:motivation_cache_hitrate}(d)).
	Consequently, having larger embedding caches provide only incremental
	improvements in further absorbing CPU memory traffic.  Conversely, for
	datasets with low locality 
	(\fig{fig:motivation_cache_hitrate}(a)), achieving $>$$90\%$ cache hit rates
	would require more than $65\%$ of the embedding table entries to be cached in
	GPU memory.  Given state-of-the-art \recsys require TB-scale memory
	capacity, caching such high proportion of embedding tables is an impossible
	design point to pursue given the ``meager'' tens of GBs of GPU memory.

\subsection{Our Goal: Training Embedding Layers at GPU Memory Speed}
\label{sect:motivation_goal}

Our characterization root-caused the memory bandwidth limited embedding
layer training  as a key limiter in \recsys
training's performance. Caching most-frequently-accessed, hot embeddings inside
high-bandwidth GPU memory could alleviate the performance penalties of sparse
embedding gathers and gradient scatters, but the limited GPU memory capacity
makes it impossible to fully capture the active working set of embedding
tables. This makes GPU embedding caches to still suffer from the latency
overhead of bringing in the missed embeddings from CPU memory, which sits on
the critical path of \recsys training and hurting performance.  
A key
	objective of our study is to develop a GPU embedding cache that
		is able to intelligently store (and evict) not just the current but also
		\emph{future} embedding table accesses such that its active working set 
		is ``always'' available within GPU memory by the time embedding layer
		training is initiated. This would allow the training of embedding
		layers to occur at the speed of high-throughput GPU memory, drastically
		improving the performance of end-to-end \recsys training. 

\section{\proposed Architecture}
\label{sect:proposed}

In \sect{sect:proposed_principle}, we first elaborate on the design objectives 
behind the proposed \proposed architecture.
\sect{sect:proposed_strawman} then discusses our naive, straw-man architecture
which we utilize to point out the important research challenges this paper
innovates in. Finally, \sect{sect:proposed_pipelined} presents our 
\proposed architecture with its implementation details discussed in 
\sect{sect:proposed_implementation}.

\begin{figure}[t!] \centering
        \includegraphics[width=0.475\textwidth]{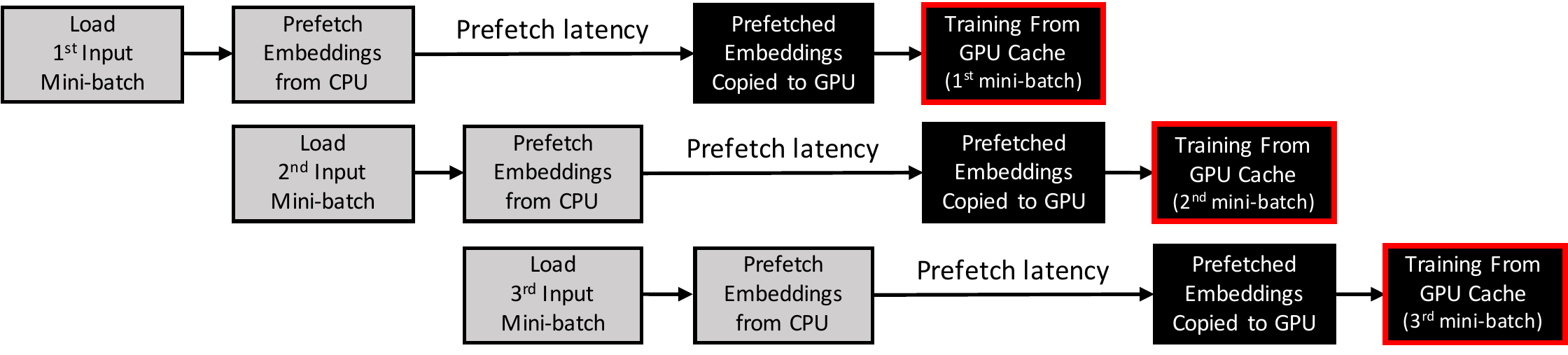}
\caption{
High-level overview of \proposed's key approach.
}
\vspace{-1.3em}
\label{fig:proposed_motivation}
\end{figure}

\subsection{Design Overview}
\label{sect:proposed_principle}

{\bf Design objectives.} \proposed is a software runtime system that manages
high-bandwidth GPU memory as a fast \emph{scratchpad} for servicing embedding
gathers and gradient scatters. Our GPU scratchpad is designed to carefully (and
		systematically) bring in the required embeddings from CPU$\rightarrow$GPU
right before the embedding training procedure is initiated, so it functions as
an embedding cache that \emph{always hits}. This allows \proposed to
	train embedding layers virtually at GPU memory speed, which is a level of
		performance only achievable when scaling out to \emph{multi-GPU} systems.
		In other words, to train embedding layers at GPU memory speed under
		conventional systems, all the embedding tables must be partitioned and
		distributed across multiple GPU's local memory so that embedding table
		accesses are always captured within the multi-GPUs' memory pool.  A key
		design objective of \proposed is to reach commensurate performance of such
		``GPU-only'' system using a \emph{single} GPU machine, thereby
		significantly reducing \recsys training cost. Later in
		\sect{sect:eval_cost_new}, we evaluate the advantages  of our \proposed vs.
		multi-GPU systems in terms of training cost reduction.

{\bf Key observations.} Conventional caches are not able to achieve the
aforementioned research objective (i.e., a cache that always hits) because the
cache insertion/replacement policy is a \emph{reactive} one based on a
best-effort speculation of what \emph{may} happen in future memory accesses
based on past history. The ``optimal'' cache design we pursue must know exactly
when and how many data elements will be accessed soon and utilize that
information to \emph{proactively} ``prefetch'' the required data into the cache
\emph{just before the on-demand cache accesses occur}, enabling them to all become
cache hits.

Our key observation is that, with respect to embedding
layer training, it is actually
possible to \emph{precisely} know when and how many embedding table accesses
will occur in the future. More concretely, the sparse  feature IDs used
for embedding gathers and gradient scatters (\fig{fig:recsys_training}) are
already recorded as part of the \emph{training dataset}. This is because the embeddings corresponding to the sparse
feature IDs will be the primary target for model updates, so the training
dataset contains the indexing information to the embedding tables,
				i.e., what memory locations to reference for embedding gathers and
				gradient scatters.  In other words, the training dataset is, by design,
				implemented to provide exactly which rows within the embedding table to
				read (write) from (to), not just for the current but also for future
				training iterations.  

				The novelty of \proposed is the utilization of such information to
				develop our GPU embedding cache architecture that can ``always''
				fully service embedding gathers and gradient scatters over GPU
				memory. \fig{fig:proposed_motivation} illustrates our key approach
				in designing \proposed. Conventional GPU embedding caches
				frequently invoke cache misses which require the missed embeddings
				to be reactively fetched from the CPU embedding tables to the GPU, causing
				latency overheads.  Note that once the missed embeddings are
				copied over to GPU memory, we are able initiate \recsys training
				at GPU memory speed. 
				In \proposed, the training dataset is examined in advance to \emph{prefetch}
				soon-to-be-accessed embeddings from CPU memory and copy them over to the GPU.
				To \emph{hide} the 
				latency overhead of CPU$\rightarrow$GPU embedding prefetching, \proposed
				collects \emph{multiple} mini-batches worth of sparse feature
				IDs from the training dataset, concurrently processing
				\emph{different} stages of multiple training iterations via
				``pipelined'' execution.  As depicted in \fig{fig:proposed_motivation},
						hiding the latency of copying prefetched embeddings from
						CPU$\rightarrow$GPU enables a single \recsys training
						iteration to be completed every single pipeline ``cycle'' (the
								red-colored stages in \fig{fig:proposed_motivation}).  A
						key challenge of such pipelined design is that the
						concurrently executing mini-batches' embedding table accesses
						can potentially interfere with each other's GPU embedding
						cache lookups, hindering the correct execution of \recsys
						training algorithm. Consequently, the research problem lies in
						how to intelligently manage the GPU embedding cache when
						multiple input mini-batches are concurrently in-flight,
						without violating the correctness of program execution.  In
						the next subsection, we present a naive, straw-man
						architecture for \proposed which we utilize as a driving
						example to discuss its flaws and limitations for realizing the
						vision behind \fig{fig:proposed_motivation}, motivating the
						important research challenges we address in this work.

\subsection{Straw-man Architecture of \proposed}
\label{sect:proposed_strawman}

{\bf The need for dynamic (not static) embedding caches.} The \emph{static}
embedding cache we assume in \fig{fig:baseline_vs_caching_flow}(b) {is always
filled in with the top-$N$ hot embeddings without eviction.  \proposed
requires the ability to \emph{dynamically} determine what embedding table
accesses will occur for the current and upcoming mini-batches and utilize that
information to 1) proactively prefetch the embeddings (currently missing in
the GPU embedding cache) from CPU memory and 2) transfer them to the GPU
embedding cache in advance, before the on-demand embedding fetches 
occur.  Designing such \emph{dynamic} embedding cache requires the following
capabilities: 1) determine which among the sparse feature IDs are hits or
misses, 2) utilize that  information to fetch the missed embeddings from the
CPU embedding tables, and 3) choose a number of embeddings (equivalent to the
	number of missed IDs) to evict from the GPU embedding cache, so that the
missed embeddings fetched from the CPU can be filled into 
the GPU embedding cache. As shown in
	\fig{fig:proposed_strawman}, our straw-man with a dynamic GPU
	embedding cache goes through the following steps for training:

\begin{figure}[t!] \centering
        \includegraphics[width=0.48\textwidth]{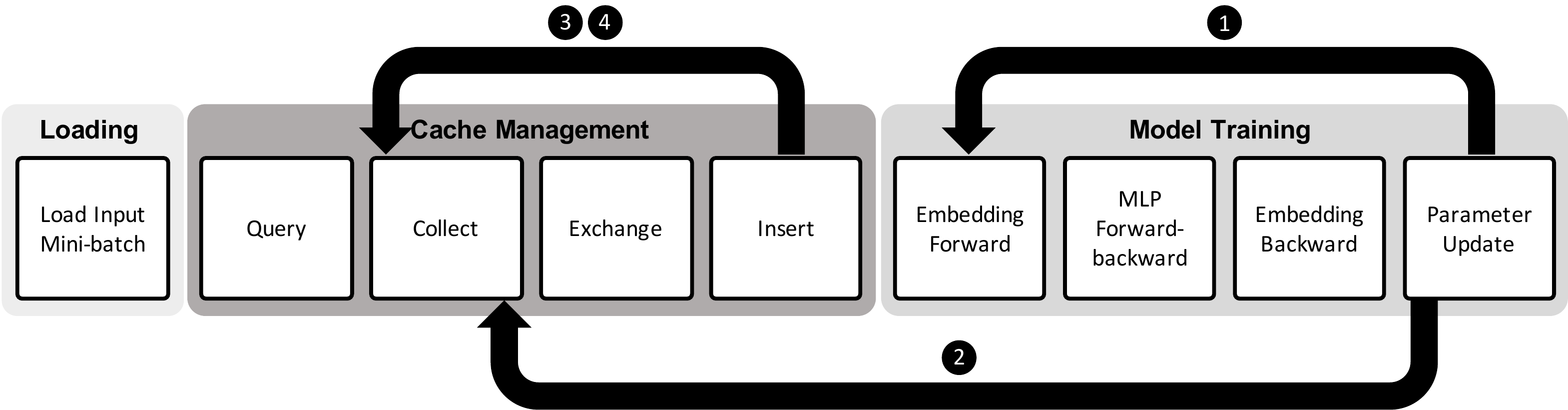}
\caption{
Key steps undertaken during straw-man architecture's single iteration of \recsys training. The black circles ($1$-$4$) refer to the RAW dependencies at the GPU embedding cache ($1$,$2$,$3$) and the CPU embedding table ($4$).
}
\vspace{-1.5em}
\label{fig:proposed_strawman}
\end{figure}

\begin{enumerate}

\item {\bf [Query]} stage: copies the sparse feature IDs for the current
training iteration from CPU$\rightarrow$GPU and query the GPU embedding cache,
				 determining the hit/miss IDs. 

\item {\bf [Collect]} stage: utilizes the missed IDs to fetch the
corresponding embeddings from CPU embedding tables. Concurrently, the GPU
fetches the same number of victim embeddings from the GPU embedding cache so
that 1) the evicted embeddings can be written backed into CPU's embedding tables,
		 and 2) the missed embeddings fetched from the CPU can be inserted into
		 the evicted entries.  The reason why the GPU embedding cache \emph{must} writeback 
		 the evicted embeddings to the CPU
		 memory is because the GPU embedding cache holds dirty copies of
		 the trained embeddings (i.e., the \emph{trained} embeddings).

\item {\bf [Exchange]} stage: copies the GPU cache missed embeddings from
CPU$\rightarrow$GPU while also simultaneously copying the evicted embeddings from
GPU$\rightarrow$CPU over PCIe.

\item {\bf [Insert]} stage: uses the CPU$\leftrightarrow$GPU exchanged
embeddings to update the CPU embedding tables with the (GPU embedding cache) evicted
embeddings while the GPU fills in the missed embeddings into the GPU embedding cache.

\end{enumerate}

Once we reach the [Insert] stage, the GPU embedding cache now holds the complete
set of embeddings required to go through forward and backpropagation of
embedding layers. Specifically, the [Embedding Forward] stage gathers all the
required embeddings from the GPU embedding cache (which will all be hits)
	and goes through the
remaining MLP forward and backpropagation. Once the final gradients are ready,
					the [Parameter Update] stage overwrites the corresponding rows in
					the GPU embedding cache with the updated model values.
Consequently, \emph{all} embeddings inserted into the (dynamic) GPU embedding cache 
are subject for training, thus \emph{any entries that get evicted from the embedding 
cache must be written back into the main CPU embedding tables}.

\begin{figure}[t!] \centering
    \includegraphics[width=0.475\textwidth]{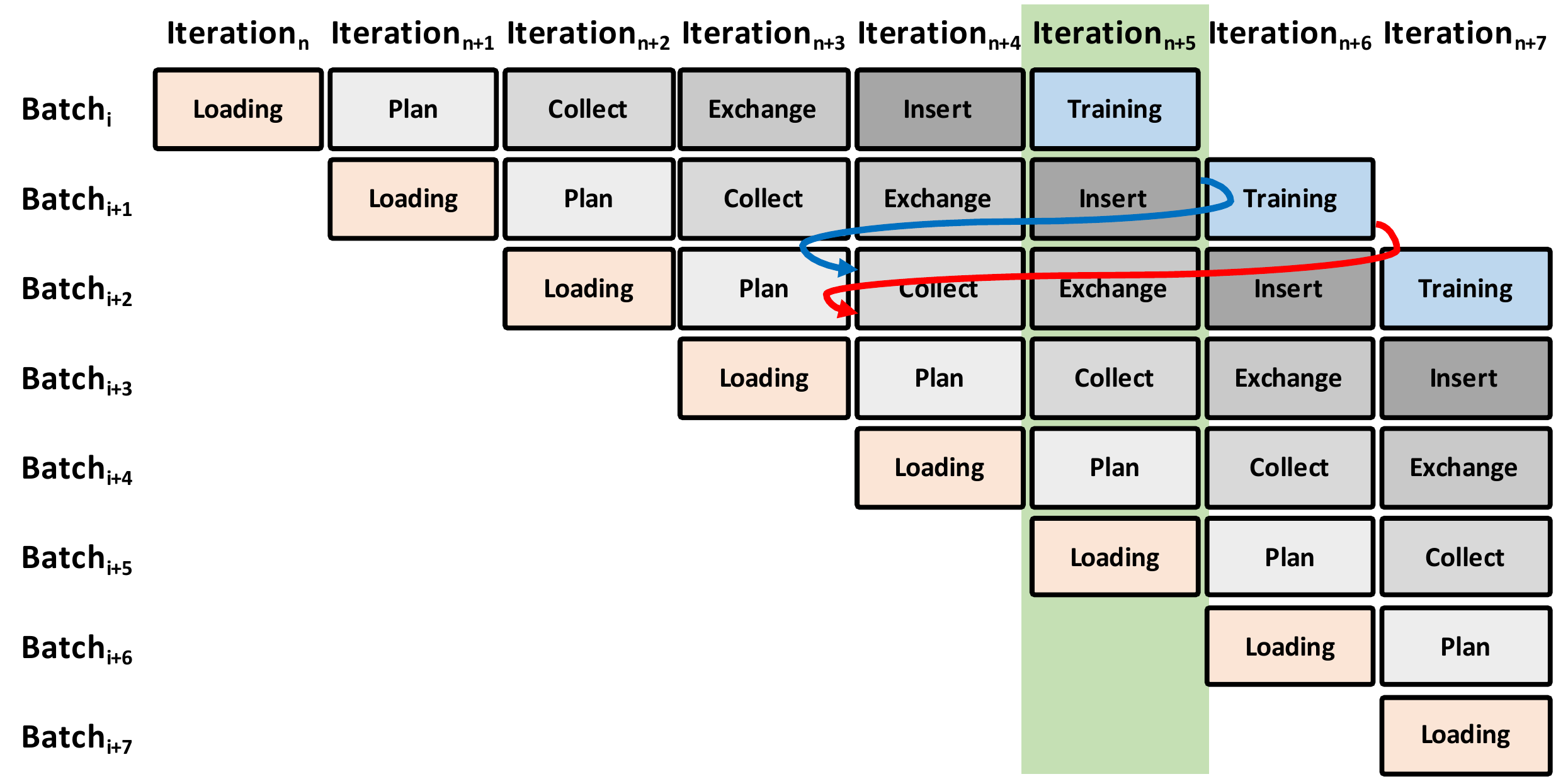}
    \caption{
        Pipelined version of straw-man architecture. The red and blue 
					arrows designate the RAW-(2) and RAW-(3)/(4)
				dependencies, respectively, causing pipeline hazards. RAW-(1) is eliminated by having
						the four steps of [Model Training] stages as a single [Training] step (detailed in \sect{sect:proposed_pipelined}). Assuming
					the data hazards are resolved however, pipelining the straw-man
					design enables a single mini-batch of training to be completed each cycle, effectively hiding the latency overheads of the [Cache Management] stages.
        }
        \vspace{-1.3em}
        \label{fig:proposed_pipelining_example}
\end{figure}

{\bf Limitations of straw-man architecture.} A critical challenge of 
straw-man is that the [Cache Management] stages sit in the critical path of
training.  As discussed in \fig{fig:proposed_motivation}, our goal is
to collect multiple training mini-batches and concurrently process different
stages of multiple training iterations via pipelined execution, which will
enable the embeddings collected from [Query$\rightarrow$Collect] stages to have
the effect of being prefetched into the GPU.  Unfortunately, the multiple
stages in straw-man invoke several \emph{read-after-write
	(RAW) data dependencies} at both the GPU embedding cache and the CPU
	embedding tables, which could prevent straw-man from concurrently executing
	different stages in \fig{fig:proposed_strawman}.  Specifically, the GPU
	embedding cache becomes a source of RAW dependencies during the following
	pair of stages:
			
\begin{itemize}

\item $[$Parameter Update] \WR$\rightarrow$ [Embedding Forward] \RD (RAW-\encircle{1} in \fig{fig:proposed_strawman}): write trained embedding values to update the embedding cache $\rightarrow$ read embeddings from the embedding cache for forward propagation.

\item $[$Parameter Update] \WR$\rightarrow$ [Collect] \RD (RAW-\encircle{2}): write trained embedding values to update the embedding cache $\rightarrow$ read victim embeddings from embedding cache.

\item $[$Insert] \WR$\rightarrow$ [Collect] \RD (RAW-\encircle{3}): write missed embeddings into embedding cache $\rightarrow$ read out victim embeddings from embedding cache.

\end{itemize}

Similarly, the CPU embedding table becomes another source of RAW dependency during [Insert] \WR 
(i.e., write evicted embeddings into CPU embedding table) and
[Collect] \RD (i.e., read missed embeddings from CPU embedding table), RAW-\encircle{4} in \fig{fig:proposed_strawman}. 
All of these RAW dependencies are naturally resolved without
any complications when different training iterations are sequentially executed. 
When we seek to pipeline the straw-man architecture however, the RAW dependencies cause a \emph{data hazard} 
inside the pipeline and prevents the correct execution of \recsys training (\fig{fig:proposed_pipelining_example}).

\subsection{``Pipelined'' \proposed Architecture}
\label{sect:proposed_pipelined}

Our final proposition for \proposed, its ``pipelined''
version, holistically addresses all the shortcomings of our straw-man
with our principled design approach.
For clarity of explanation, we henceforth refer to the pair
of stages that incur RAW dependencies (\encircle{1} to \encircle{4}) as
\emph{RAW dependent stages}.

The pipelined \proposed
incorporates the {\bf [Plan]} stage (as a substitute for [Query] in
		straw-man) in its $6$-stage pipeline
(\fig{fig:proposed_pipelined}), which provides
the following functionality:

\begin{enumerate}
\item The [Plan] stage is
the main control unit that \emph{plans out} in advance which embeddings to
gather and fill (from CPU embedding tables to GPU embedding cache) and evict
and write-back (from GPU embedding cache to CPU embedding tables) during the
remaining pipeline stages, specifically during [Collect] and [Insert].

\item All the procedures undertaken during the straw-man's [Query]
stage are also conducted in [Plan], i.e., copying the sparse feature IDs
for the current mini-batch from CPU$\rightarrow$GPU and determining
the hit/missed IDs by querying the GPU embedding cache.
\end{enumerate}

Overall, the goal of \proposed's [Plan] stage is to 
remove any RAW hazards from occurring inside the pipeline so that
\proposed's prefetching latency is effectively hidden without causing
any complications to \recsys training.

\begin{figure}[t!] \centering
        \includegraphics[width=0.45\textwidth]{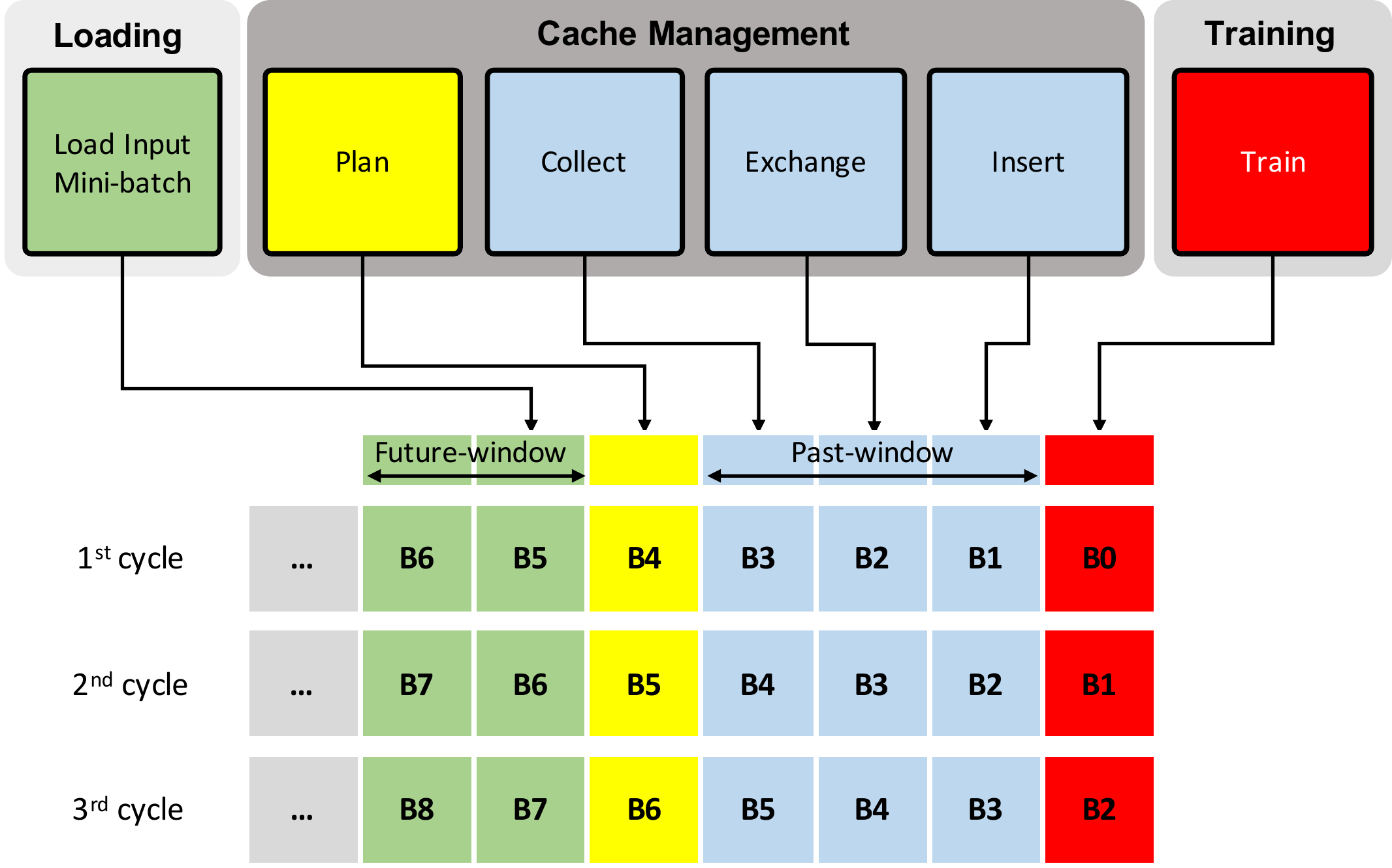}
\caption{
	Pipelined \proposed architecture. 
	The [Plan] stage examines six mini-batches (three past, one current, and two future) that fall under 
	the current (sliding) window to determine GPU embedding cache victims that do not cause RAW hazards. 
	B($N$) refers to $N$-th input batch.
}
\vspace{-1.3em}
\label{fig:proposed_pipelined}
\end{figure}

{\bf RAW dependencies in ``sparse'' embedding tables.}
Recall that a unique property of embedding table accesses is that they are
extremely sparse, i.e., a single training iteration only touches upon several
tens of thousands of rows within the several millions to billions of entries
in the embedding tables. The implication of such property from a RAW
data dependency perspective is as follows: as long as the row IDs used for the
reads/writes to/from the GPU embedding cache and the CPU embedding tables do
not overlap across RAW dependent stages, we are able to completely remove the
\emph{effectual} data dependencies, thus resolving any data \emph{hazards} from
occurring. The key is to make sure that the set of row IDs
used for ``read'' operations \RD from GPU embedding cache or CPU embedding
table do not match the row IDs used for ``write'' operations \WR across
RAW dependent stages. Below we elaborate on our principled approach to
eliminate \emph{all} RAW dependencies in our pipelined \recsys training.

{\bf Handling RAW-\encircle{1}.} Because the sparse feature IDs used in
[Embedding Forward] exactly match the IDs used for [Parameter Update], its
RAW dependency is a fundamental one that cannot be removed and must be
honored in all circumstances.  As shown in
\fig{fig:proposed_pipelined},  \proposed handles this RAW dependency
by executing all four steps of [Model Training] as a single \emph{stage}
(denoted as the [Training] stage) within \proposed's pipelined execution,
	honoring its dependency. 

{\bf Removing RAW-\encircle{2}/\encircle{3}.} To prevent
RAW-\encircle{2}/\encircle{3} from causing data hazards, \proposed should
guarantee that the GPU embedding cache entries scheduled for updates through
writes (i.e., updated during [Parameter Update] or [Insert] \WR) are never
prematurely chosen as victims to be read out from the embedding cache and
written back into CPU embedding tables. \proposed handles both
RAW-\encircle{2}/\encircle{3} dependencies through the following mechanism:
\emph{when [Plan] chooses victims to evict out of the GPU embedding cache, \proposed
	does not consider those set of input sparse feature IDs used during the
		previous $three$ training iterations (i.e., the distance between the
				[Training] stage and the [Collect] stage) as victims.}  
		At steady-state, a total of six 
		training mini-batches (corresponding to six sets of input sparse feature
				IDs) are concurrently being processed -- but executing at different
		stages.  When an input mini-batch enters the [Plan] stage, the
		\proposed controller examines three previous mini-batches (executing in
				[Collect-Exchange-Insert] stages from the [Plan] stage's
				perspective) and rules out all of their input sparse feature IDs from
		embedding cache eviction candidates. This allows the input mini-batch
		currently executing in [Plan] to never evict (i.e., read \RD)
		any one of the embeddings that will be updated (\WR) by previous input mini-batches executing in RAW dependent stages, effectively eliminating
		hazards.

{\bf Removing RAW-\encircle{4}.} Preventing hazards for
RAW-\encircle{2}/\encircle{3} is a matter of \emph{controlling}
the ``read'' part of the RAW dependency (i.e., ``writes'' to GPU embedding
		caches are done by \emph{previous} input mini-batches which the [Plan]
		stage has no control over, so \proposed controller judiciously rules out
		potentially hazard invoking row IDs from eviction to prevent RAW-\encircle{2}/\encircle{3}).
Removing RAW-\encircle{4} is the opposite as the source of RAW dependency is at
the CPU embedding tables. Here ``writes'' to the CPU embedding tables is an
artifact of the [Plan] stage choosing GPU embedding cache victims. Because
the evicted entry's write-back to embedding tables occurs during the
[Insert] stage, preventing RAW hazards for RAW-\encircle{4} is a matter of
\emph{controlling} the ``write'' part of the RAW dependency and making sure
\emph{upcoming}, future input mini-batches do not conflict with the currently
chosen eviction candidates (e.g., from the perspective of the mini-batch
		executing in [Insert], the mini-batch executing in the RAW dependent
		[Collect] stage is a \emph{future} input). \proposed removes
RAW-\encircle{4} via the following mechanism: 
\emph{when [Plan] chooses victims to evict out of the GPU embedding cache, \proposed
	does not consider those set of input sparse feature IDs used during the
		next two upcoming training iterations (i.e., the distance between
				[Insert] and [Collect]) as victims.}

{\bf Putting everything together.} As depicted in \fig{fig:proposed_pipelined},
	\proposed systematically evaluates potential RAW hazards and removes them
via our sliding window based [Plan] stage. When an input mini-batch enters
the [Plan] stage, the three previous (past-window) as well as the next two
(future-window) mini-batches of sparse feature IDs are examined. 
By creating a superset of the sparse feature IDs
falling under the past-/future windows, the [Plans] control unit
rules out the IDs included in the superset from cache eviction candidates,
			preventing RAW hazards from occurring in 
subsequent stages. 

\begin{figure}[t!] \centering
    \includegraphics[width=0.478\textwidth]{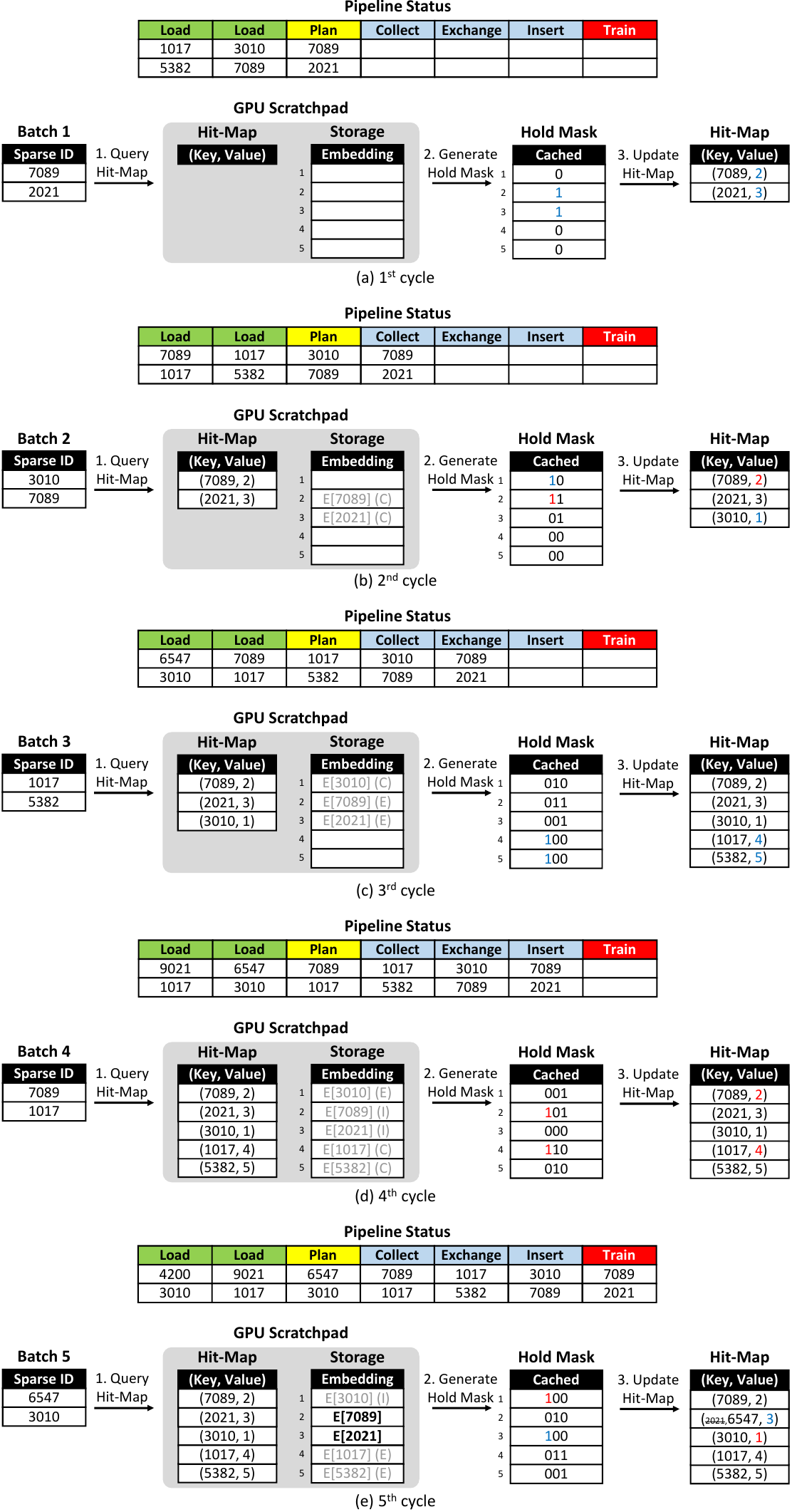}
\caption{
Example showing key data structures used in \proposed's design.  The gray-colored E[$N$]($S$) inside the Storage array designates an embedding located at the row ID $N$, scheduled for a \emph{fill} operation to Storage, is propagating through the pipeline at stage $S$ (C=[Collect], E=[Exchange], I=[Insert]). For ease of explanation, the Hold mask only shows the three bitmasks tracking the three mini-batches falling under the past-window (no future-window). The red (blue) `$1$'s in the Hold mask means a hit (miss) occurred at the Hit-Map.
}
\vspace{-1.3em}
\label{fig:proposed_hitmap}
\end{figure}

\subsection{Implementation}
\label{sect:proposed_implementation}

We now discuss \proposed's implementation details using
\fig{fig:proposed_hitmap} as an illustrative example.  
\algo{algo:scratchpipe}
provides a pseudo-code of the key operations
	conducted by ScratchPipe's controller
over its key data structures. For brevity and consistency
with
\fig{fig:proposed_hitmap}, \algo{algo:scratchpipe} 
assumes the Hold mask only holds three mini-batches
	falling under the past-window.

{\bf GPU scratchpad design.} As depicted, \proposed's
GPU embedding cache (i.e., the GPU scratchpad) is implemented using 1) a data array
to store the cached embedding vectors (denoted {\bf Storage}), and 2) a (key,
	value) store called {\bf Hit-Map} which returns the cache query results as a
hit or a miss, i.e., the (key, value) stores the cached embedding's original 
sparse feature
ID (key) and the index to locate the cached embedding within the Storage
array (value).  Whenever any given mini-batch enters the [Plan] stage, the GPU
scratchpad's Hit-Map is queried to derive the hits/misses, the result of which
is used to schedule which embeddings to gather from the CPU embedding table (if
		any) and which entries to evict from the GPU scratchpad (if any).  A unique
property of our GPU scratchpad design is that the status of the Hit-Map and Storage is
updated in a (purposefully) \emph{asynchronous} and \emph{delayed} manner.
As depicted in \fig{fig:proposed_hitmap}, the status of the Hit-Map is updated 
whenever a new mini-batch is processed at [Plan], whereas the Storage
array gets updated when a mini-batch enters [Insert] with a Hit-Map miss.
This is an artifact of \proposed's pipelined design, rendering
the status of Hit-Map to always reflect the embedding caching status of the Storage
array ``four'' cycles later in the future (i.e., the distance between [Train]
		and [Plan]). For instance, the second mini-batch of ID $3010$/$7089$
is returned as miss/hit when querying the Hit-Map during the 2nd cycle,
		even though the Storage array is still left vacant (\fig{fig:proposed_hitmap}(b)).
Such discrepancy between the status of Hit-Map and Storage is intentional because
1) the Storage array \emph{should not}
have yet to be updated with
the first mini-batch's query of ID $7089$/$2021$ as it has not arrived
at the [Insert] stage, and 2) the second mini-batch should still be
able to \emph{see} the precise caching status of GPU scratchpad (i.e., the
status assuming the first
mini-batch completed its training) so that it can accurately determine the set
of embeddings to prefetch from CPU, even though
it is currently at [Plan] stage.

\begin{algorithm}[t!]
\scriptsize
	\caption{ScratchPipe Pipeline Controller}
	\label{algo:scratchpipe}
	\begin{algorithmic}[1]
	\LineComment{Key data structures for ScratchPipe control}
	\State HitMap = [(key,value) storage to index cached embeddings in storage]
	\State HoldMask = [circular queue tracking cached locations]
	\State $HoldMaskWidth$ $\leftarrow$ 3
	\Statex

	\LineComment{Step A: A new mini-batch enters the [Plan] stage}
	\State MiniBatch $\leftarrow$ PipelineMiniBatchQueue["Plan"]
	\Statex

	\LineComment{Step B: Advance HoldMask by one cycle}
	\For {$i$ $\leftarrow$ $1$ to $CacheSize$}
	    \State HoldMask[$i$] $\leftarrow$ HoldMask[$i$] $>>$ $1$
	\EndFor
	\Statex

	\LineComment{Step C: Iterate through all sparse IDs in mini-batch to determine hits and misses}
	\For {$i$ $\leftarrow$ $1$ to $NumberOfSparseIDsWithinMiniBatch$}
		\LineComment{If a sparse ID is found in Hit-Map, set the corresponding HoldMask}
	    \If {MiniBatch[$i$] is found in HitMap}
	        \State $hitIdx$ $\leftarrow$ HitMap[MiniBatch[$i$]]
			\State HoldMask[$hitIdx$] $\leftarrow$ HoldMask[$hitIdx$] $|$ $2^{HoldMaskWidth-1}$
			\State \hskip-1.5em \(\triangleright\) If sparse ID misses, select victim index whose HoldMask is "$0$"
		\Else
		    \State $evictIdx$ $\leftarrow$ \Call{ChooseVictim}{HoldMask}
		    \State HitMap[MiniBatch[$i$]] $\leftarrow$ $evictIdx$
			\State HoldMask[$evictIdx$] \hskip-0.15em $\leftarrow$ \hskip-0.15em HoldMask[$evictIdx$] $|$ $2^{HoldMaskWidth-1}$
		\EndIf
	\EndFor
	\end{algorithmic}
\end{algorithm}

{\bf Hold masks for removing RAW hazards.} \proposed removes RAW hazards by
having the [Plan] stage examine the three previous (past-window) and two next
(future-window) mini-batches' embedding table look-up IDs falling under the
current sliding window (centered around [Plan]), making sure those cached
locations are not evicted from the GPU scratchpad (\sect{sect:proposed_pipelined}). As shown in
\fig{fig:proposed_hitmap}, we employ a data structure named {\bf Hold
	mask}, which is a bitmask (number of bits within the bitmask is equal to the number of
	cacheable slots in Storage array)
	the [Plan] stage utilizes to designate locations within the
	scratchpad's Storage array that the current mini-batch will be utilizing at
	the [Train] stage. The Hold mask is designed using a circular
	queue to accommodate its sliding window based operation,
	storing
	a set of 
	six bitmasks 
 that keep track of the past three,
	one current, and two future mini-batches' cached locations in Storage -- ones
	that should \emph{not} be targeted for eviction while the sliding window
	is effectual. 
When the control unit in [Plan] stage needs to select a victim
	for GPU scratchpad eviction, the Hold mask is examined.
	The victim candidates are chosen as the Storage array locations
	that correspond to the Hold mask's bit-locations set with a value of '0'.
	This means
			none of the sparse feature IDs that fall under the current
			sliding window
			asked to \emph{hold} this location inside the GPU scratchpad, hence is
			allowed to be evicted. 			For instance,
			E[2021] stored as the 3rd element inside the Storage array is targeted for eviction
			at the 5th cycle as the corresponding location in the Hold mask is ``000''
			after the 4th cycle (\fig{fig:proposed_hitmap}(d,e)).

	It is worth pointing out that, in order for our \proposed architecture
	to guarantee that the
	bit-locations within the Hold mask set to non-zero values are never targeted
	for scratchpad eviction, the Storage array should be large enough to
	accommodate the worst-case GPU scratchpad usage within the processing of
	six mini-batches falling under the current window.  We quantify the
	implementation overhead of our GPU scratchpad
	in~\sect{sect:eval_overhead}.  Furthermore, \sect{sect:sensitivity}
	evaluates \proposed  sensitivity to model configurations, replacement policies, etc. 

\section{Methodology}
\label{sect:methodology}

{\bf Hardware.} \proposed is evaluated on a server 
containing Intel Xeon E5-2698v4 ($256$ GB DDR4, $76.8$ GB/sec DRAM bandwidth) and
NVIDIA's V100 ($32$ GB, $900$ GB/sec of DRAM bandwidth).
The CPU and GPU communicates over PCIe(gen3) with $16$ GB/sec of communication bandwidth.

{\bf Software.} Our software runtime system is designed using a combination of PyTorch
(v1.8.0)~\cite{torch} for modeling the training dataset loader and the
embedding layers with NVIDIA's cuBLAS/cuDNN~\cite{cublas,cudnn} for modeling
DNN layers.  Since \proposed is implemented purely in software, we measure
end-to-end wall clock time when reporting performance.

{\bf Benchmarks.} Because real-world traces used for \recsys training are not
publicly available, we generate a synthetic embedding table access trace as
follows.  As discussed in \sect{sect:motivation_locality}, the locality
inherent in \recsys datasets varies significantly depending on its application
domain. To properly reflect the diverse locality characteristics of different
\recsys models, the sorted access count of embedding table entries in various
real-world datasets (\fig{fig:motivation_locality}) are utilized to generate a
wide ranging set of probability density functions (PDFs) that quantifies an
embedding table entry's likelihood of lookup.  Depending on the dataset, some
PDFs exhibit low locality (e.g., User table in Alibaba~\cite{alibaba_taobao})
	while others exhibiting medium, or high locality (e.g.,
			Criteo~\cite{criteo:terabyte}).  To properly capture a wide range of \recsys
	embedding table access patterns in our evaluation, we use these PDFs to generate
	three types of embedding table access traces, exhibiting low, medium,
		 and high locality.
We additionally add a random trace (i.e., embedding table access IDs are randomly
		generated) to compare against low, medium, and high locality traces.
		 These input traces are fed into our
		baseline \recsys model to generate four distinct benchmarks to stress test
		the performance of our evaluated cache architectures. The baseline \recsys model
		configuration is established using DLRM MLPerf~\cite{dlrm:mlperf} as well
		as representative \recsys models published by prior
		work~\cite{tcasting,mudigere2021high,facebook_dlrm},
		which has eight embedding tables, each table containing ten million
		embedding vector entries (each embedding sized as a $128$-dimensional
				vector), amounting to $40$ GB of total model size. The default \recsys
		model conducts $20$ embedding gathers per each table with a batch size of
		$2048$. In \sect{sect:sensitivity}, we study the sensitivity of
		\proposed when deviating from these default configurations.

{\bf Embedding cache size.} The size of GPU-side embedding cache has paramount
effect on overall performance. Since high-end server class GPUs typically come
with several tens of GBs, the GPU embedding cache can practically house $<10\%$
of the several hundreds to thousands of GB scale \recsys models. We follow the
methodology adopted by prior work (e.g., the static embedding cache proposed by
		Yin et al.~\cite{ttrec} study $0.01$$-$$10\%$ caching of CPU-side embedding
		tables) and study \proposed's effectiveness across a range of
$2-10\%$ cache size.

\section {Evaluation} 
\label{sect:evaluation}

We explore four design points: two baseline architectures, 1) CPU-GPU
without caching, 2) a CPU-GPU with static GPU embedding cache, and
two design points from our proposal, 
3) our straw-man architecture \emph{without} pipelining,
	and 4) \proposed \emph{with} pipelining. Note that \proposed
	\emph{does not change the algorithmic properties of
	stochastic gradient descent (SGD) training} so the total training iterations required
		to reach a given target CTR accuracy is identical between baseline
		and \proposed.

\subsection{Latency Breakdown}
\label{sect:eval_latency_breakdown}

Before discussing end-to-end performance of \proposed, we analyze its efficacy
in addressing baseline architecture's bottlenecks.
\fig{fig:eval_latency_breakdown} shows a latency breakdown of our studied
workloads.  As discussed in \sect{sect:motivation_caching}, our two baselines
spend significant fraction of training time in conducting the memory bound
embedding layer training at the low throughput CPU memory. While storing hot
entries in GPU's embedding cache does help alleviate the bottlenecks of
CPU-side training (with larger caches generally helping reduce CPU-side
		training time), it fails to overcome the fundamental limitations of slow
CPU memory as any embedding table accesses that miss in the GPU cache must be
routed to CPU memory to conduct not just the embedding gathers (forward
		propagation) but more importantly the highly memory bound gradient
duplication/coalescing followed by scatter updates during backpropagation
(\fig{fig:recsys_training}(b), \fig{fig:baseline_vs_caching_flow}(b)).  As
shown in \fig{fig:eval_latency_breakdown}(b), \proposed substantially reduces
latency because the memory intensive embedding layer training is always
conducted over our high-bandwidth GPU scratchpad. In \proposed, the only time
CPU memory is accessed is when the embeddings to be prefetched into GPU
scratchpad are [Collect]ed and when the updated models are evicted and
[Insert]ed into CPU embedding tables. While these two stages do incur
high latency within \proposed's cache management step, the absolute time
spent in interacting with the CPU memory is significantly reduced compared to
the baseline static caching, all thanks to
our GPU scratchpad.
		
\subsection{Performance}
\label{sect:eval_perf}

\begin{figure}[t!] \centering
    \subfloat[]
    {
        \includegraphics[width=0.465\textwidth]{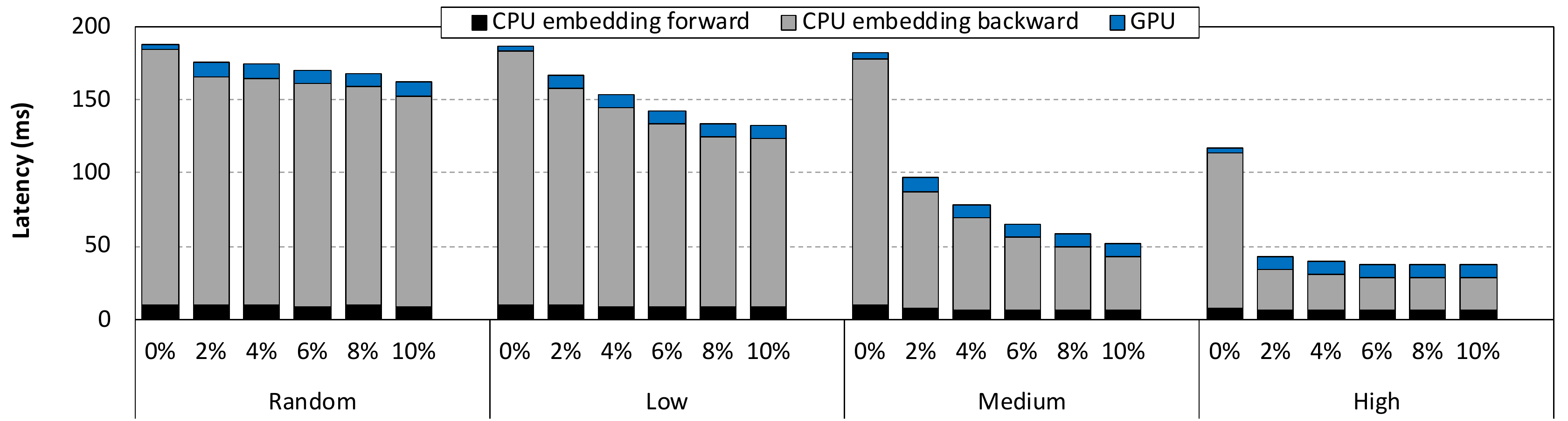}
    }
    \vspace{-0em}
    \subfloat[]
    {
    \includegraphics[width=0.465\textwidth]{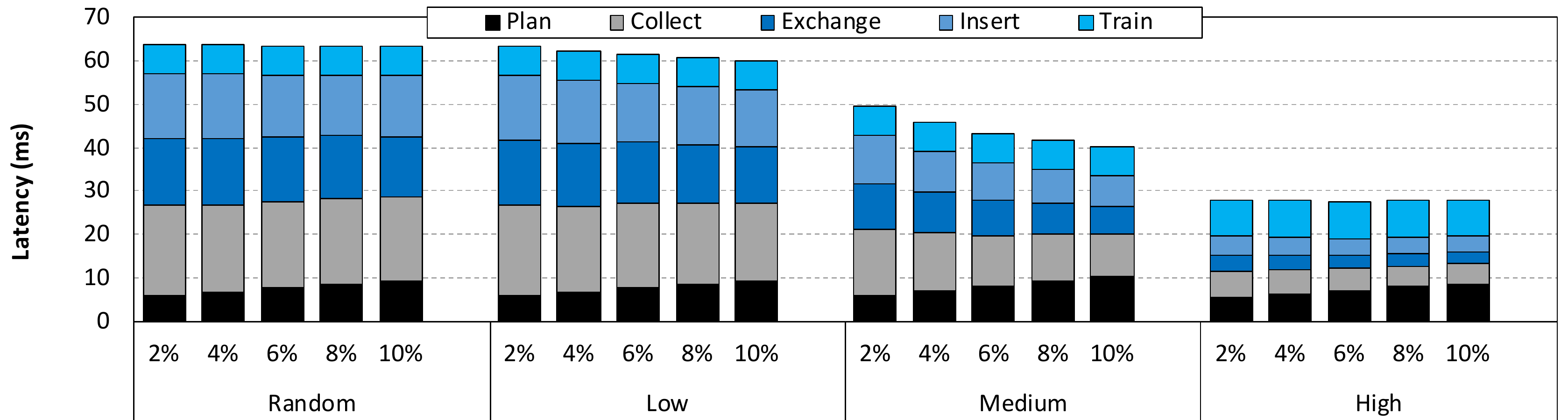}
    }
    \vspace{0em}
    \caption{
		Latency breakdown of (a) baseline CPU-GPU without caching ($0\%$) and with static GPU
			embedding cache (sized to accommodate top $2-10\%$ of hot entries in CPU embedding tables). 
			In (b), \proposed is broken down into its per-stage pipeline latency. Note that the scale
			in (a) and (b) are different ($0-200$ ms in (a) vs. $0-70$ ms in (b)).
    }
    \label{fig:eval_latency_breakdown}
		\vspace{-1.3em}
\end{figure}

\fig{fig:eval_speedup} shows \proposed's  end-to-end speedup. Across all
the data points we explore, \proposed achieves an average $2.8\times$ (max $4.2\times$)
	speedup vs. static caching, with the magnitude of performance improvement gradually
	reducing as input traces exhibit higher locality. This is expected because the more locality
	a given dataset contains, the lightweight static embedding cache design could cost-effectively
	absorb the majority of embedding table accesses. Nonetheless, even under a high locality
	workload scenario, \proposed achieves $1.6-1.9\times$ speedup, demonstrating its robustness.
	Interestingly, our straw-man without pipelining
	also provides meaningful speedup vs. static caching as it still helps reduce the number of
	gradient scatters to the CPU embedding tables, highlighting the benefits of dynamic caching vs.
	static. 

\subsection{Energy-Efficiency}
\label{sect:eval_energy}

\proposed is a purely software-based architecture
demonstrated over existing hardware/software stack.  To
measure system-level power consumption, we utilize \texttt{pcm-power}~\cite{pcm_power} for CPU's
socket-level power consumption and NVIDIA's \texttt{nvidia-smi}~\cite{nvidia_smi} for GPU power
consumption. The aggregated power is multiplied with execution time to derive energy
consumption. As depicted in \fig{fig:eval_energy}, 
\proposed's significant training time reduction directly
	translates into energy-efficiency improvements.

	\begin{figure}[t!] \centering
    \includegraphics[width=0.47\textwidth]{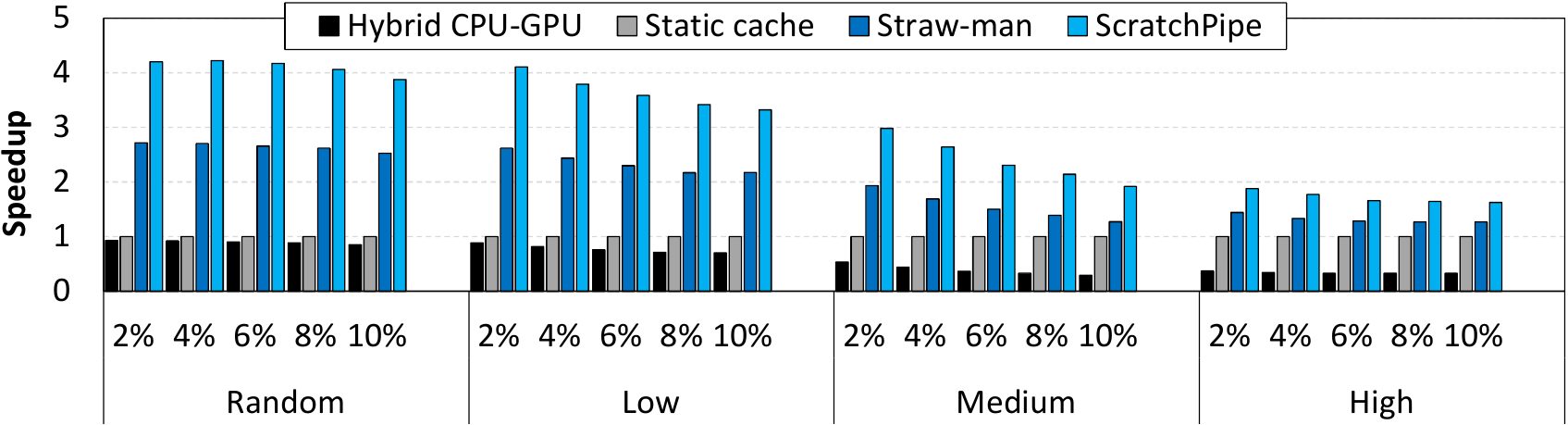}
    \caption{
				End-to-end performance speedup (normalized to static cache).
        }
        \vspace{-0.5em}
        \label{fig:eval_speedup}
\end{figure}

\begin{figure}[t!] \centering
    \includegraphics[width=0.47\textwidth]{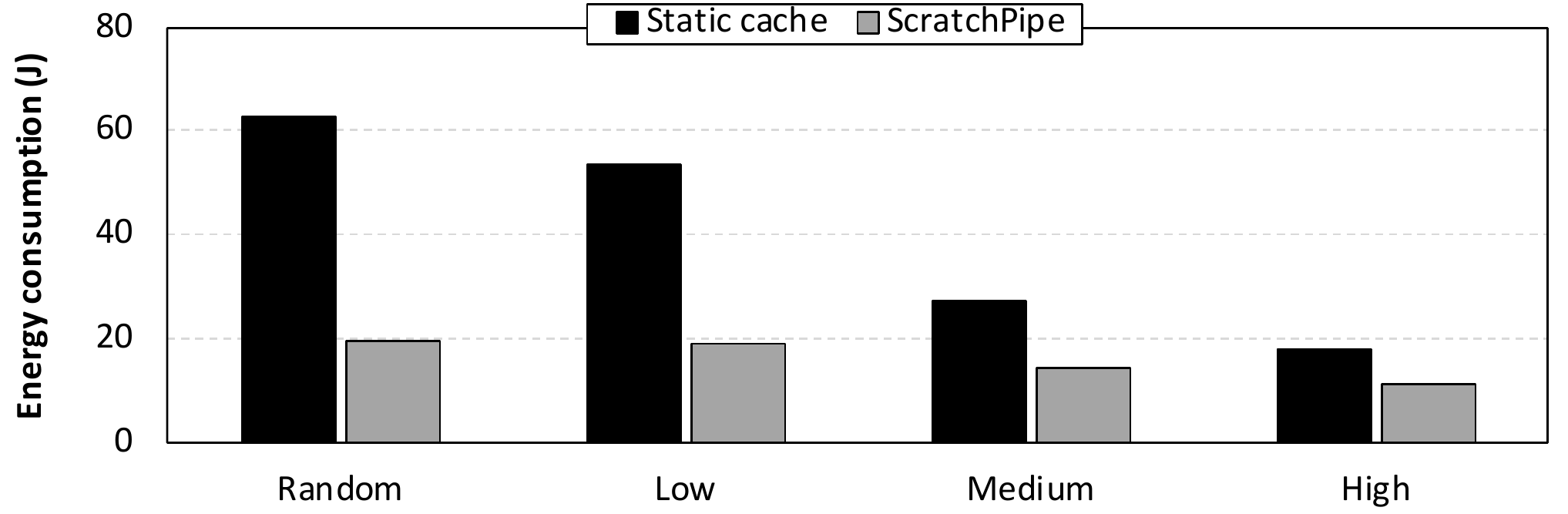}
    \caption{
				Energy consumption of static cache vs. \proposed.
        }
        \vspace{-1.3em}
        \label{fig:eval_energy}
\end{figure}

\subsection{Implementation Overhead}
\label{sect:eval_overhead}

As discussed in \sect{sect:proposed_implementation},
\proposed requires the Storage array within our GPU scratchpad to at least be
large enough to accommodate all the embeddings gathered across the six input mini-batches
currently falling under the sliding window. Under our default \recsys model
configuration, in the worst case where none of the IDs within the sliding window overlap with each other, this amounts to (number of tables $\times$ number of gathers per table
		$\times$ mini-batch size $\times$ embedding vector size) $\times$ (number of concurrent mini-batches) = ($8\times20\times2048\times128\times4$ Bytes)$\times$$6$ = $960$ MB of Storage space
provisioned for holding not-to-be-evicted embeddings. In reality however, the \emph{active} 
working set to actually hold is significantly smaller as many of the IDs gathered within
the sliding window become hits.
				In addition, the Hit-Map also incurs an additional $<$$1$ GB of memory space with other 
				miscellaneous data structures accounting for $<$$300$ MB, amounting to 
				an aggregate $<4$ GB of additional GPU side memory allocations under \proposed.

\begin{figure}[t!] \centering
    \subfloat[]
    {
        \includegraphics[width=0.475\textwidth]{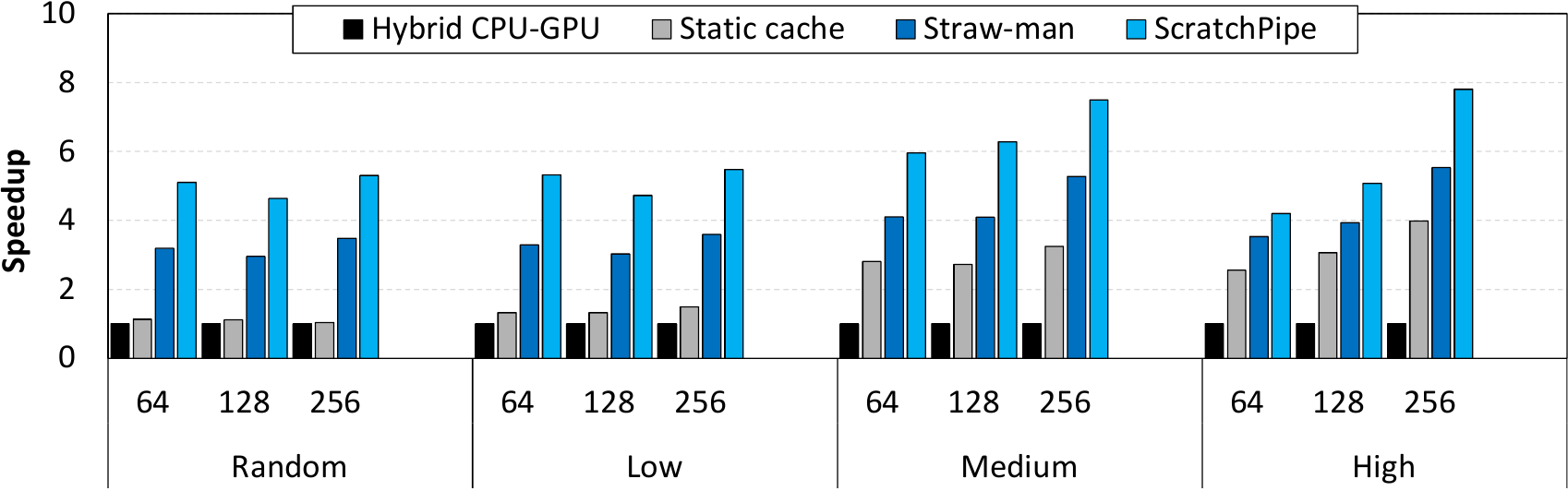}
    }
    \vspace{-0em}
    \subfloat[]
    {
    \includegraphics[width=0.475\textwidth]{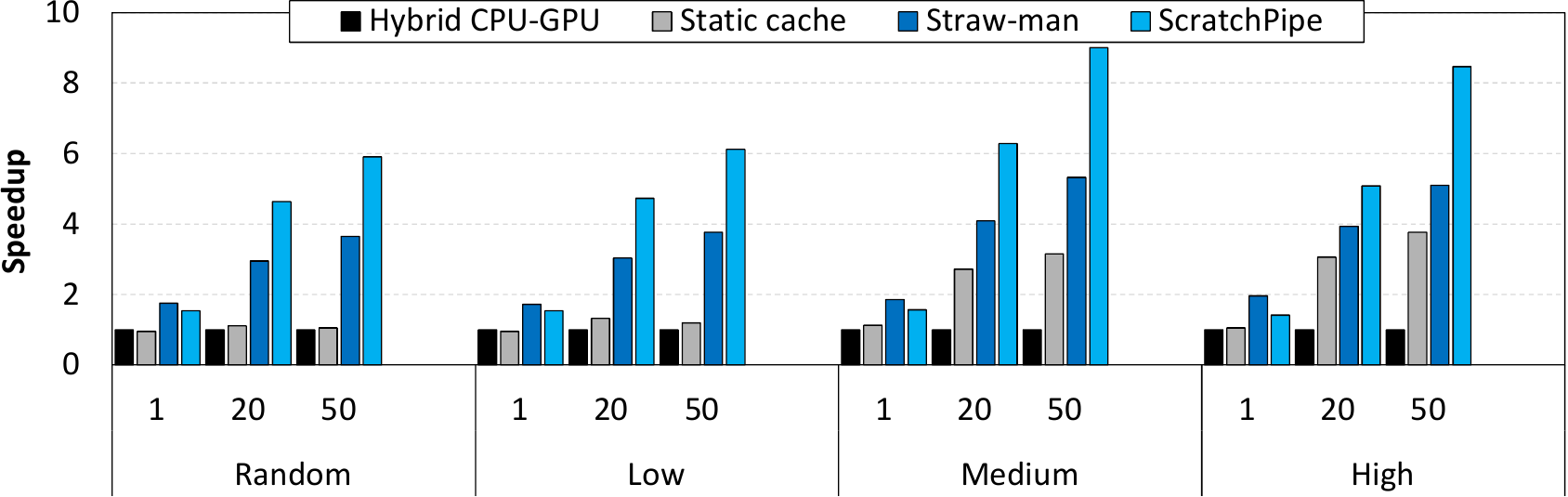}
    }
    \vspace{-0em}
    \caption{
		\proposed sensitivity to (a) embedding vector dimension size ($64$/$128$/$256$), (b) number of embedding table lookups ($1$/$20$/$50$).
    }
    \label{fig:eval_sensitivity}
		\vspace{-1.3em}
\end{figure}

\subsection{Sensitivity}
\label{sect:sensitivity}

 This section evaluates \proposed's  robustness to
\recsys model configuration or design parameters
 of \proposed.

{\bf Embedding vector dimension.} Although our default embedding vector
dimension is sized at $128$, several prior work employs smaller~\cite{recnmp}
or larger~\cite{mudigere2021high,youtube_recsys} dimension sizes in their \recsys. \fig{fig:eval_sensitivity}(a)
shows \proposed's speedup when sweeping the embedding dimension size from
$64$ to $256$. In general, the performance benefits of \proposed remains intact
across embedding dimension size with larger speedups achieved when the dimension
size gets larger. This is because larger embeddings incur an even higher memory
bandwidth pressure, rendering baseline to suffer more while \proposed's scratchpad
providing higher performance benefits.

{\bf Number of embedding table lookups.} 
\fig{fig:eval_sensitivity}(b) summarizes
the speedup \proposed achieves when the number of
embedding table gathers are swept from $1$ to $50$. 
As the number of lookups are increased to $50$, the embedding layer
causes an even more severe performance bottlenecks due to its increase
in memory traffic. Consequently, \proposed
achieves an even higher speedup with an average $3.7\times$ (max $5.6\times$). At the other end of the spectrum, we experiment an embedding table 
lookup size of $1$ which makes the \recsys model suffer less from the embedding layers. \proposed still performs better than our two baselines albeit with less improvements in performance. 
Generally, \proposed is shown to be highly robust 
across a wide range of embedding table lookup sizes.

{\bf Cache replacement policy, MLP-intensive \recsys models, etc.} In addition
to the aforementioned sensitivity studies, we also tested \proposed's
robustness under various other design points: 1) changing the GPU scratchpad's
replacement policy from our default LRU (least-recently-used) policy to a
random eviction or LFU (least-frequently-used) policy, 2) evaluating the
effectiveness of \proposed under more MLP-intensive (and less embedding
		intensive) models, 3) training under larger or smaller batch sizes, etc.
		In general, we confirm \proposed's robustness across a wide range of
		sensitivity studies. We omit the results for brevity.

\subsection{Training Cost Reduction vs. Multi-GPUs}
\label{sect:eval_cost_new}

We now discuss \proposed's benefits in reducing \recsys training cost.
	Training cost estimates are based on AWS EC2 pricing for P3 instances
		containing NVIDIA V100 GPUs (\tab{tab:tco}). \proposed does not
		change the training algorithm so the total number of training iterations
		required to reach the same level of \recsys algorithmic accuracy are
		identical between \proposed and the $8$ GPU system.  As such, we evaluate
		the training costs for the evaluated systems over $1$ million training
		iterations.  The $8$ multi-GPU  system partitions the embedding tables
		across $8$ GPUs' HBM for model-parallel training of embeddings while the
		backend MLP layers are trained using data-parallelism, enabling end-to-end
		``GPU-only'' training.  Despite using $8$ GPUs, however, the multi-GPU
		system can only reduce $29\%$/$40\%$/$64\%$/$66\%$ of training time vs.
		\proposed for high/medium/low/random datasets.  Consequently, \proposed
		consistently achieves significant training cost reduction with an average
		$4.0\times$ (maximum $5.7\times$) savings, with more cost
		savings achieved with higher locality.

\begin{table}[t!]
  \caption{Training cost of ScratchPipe vs. multi-GPU with $8$ V100 GPUs. 
  }
  \centering
\footnotesize
\scalebox{0.82}{
  \begin{tabular}{|c|c|c|c|c|c|}
    \hline
    \textbf{Dataset} & \textbf{System} & \textbf{AWS Instance} & \textbf{Price/hr} & \textbf{Iter. Time} & \textbf{1M Iter. Cost} \\ \hline
      \hline\hline
      \multirow{2}{*}{Random} & ScratchPipe   &   p3.2xlarge   & \$ 3.06  & 47.82 ms & \textbf{\$ 40.64} \\  \cline{2-6}
                              & 8 GPU         &   p3.16xlarge  & \$ 24.48 & 16.22 ms & \$ 110.3  \\  \hline\hline
      \multirow{2}{*}{Low}    & ScratchPipe   &   p3.2xlarge   & \$ 3.06  & 44.70 ms & \textbf{\$ 37.99}  \\  \cline{2-6}
                              & 8 GPU         &   p3.16xlarge  & \$ 24.48 & 16.12 ms & \$ 110.2 \\  \hline\hline
      \multirow{2}{*}{Medium} & ScratchPipe   &   p3.2xlarge   & \$ 3.06  & 29.68 ms & \textbf{\$ 25.23} \\  \cline{2-6}
                              & 8 GPU         &   p3.16xlarge  & \$ 24.48 & 17.82 ms & \$ 121.2 \\  \hline\hline
      \multirow{2}{*}{High}   & ScratchPipe   &   p3.2xlarge   & \$ 3.06  & 26.34 ms & \textbf{\$ 22.39} \\  \cline{2-6}
                              & 8 GPU         &   p3.16xlarge  & \$ 24.48 & 18.61 ms & \$ 126.6 \\  \hline
  
  \end{tabular}
}
\vspace{-1.5em}
\label{tab:tco}
\end{table}

\subsection{Discussion}
\label{sect:eval_scalability}

{\bf Applicability of single CPU-GPU ScratchPipe.}
Recent literature from Facebook elaborate on the scale of their deployed
	RecSys model size, exhibiting upto $100\times$ difference in number of
		embedding tables per model~\cite{mudigere2021software,deeprecsys}.   Specifically, RecSys models for
			content filtering are known to be smaller sized than those for
				ranking~\cite{dlrm:arch}, enabling the overall memory footprint to fit
					within a single-node's CPU memory capacity.  Given content filtering's
						wide applicability across various web-based applications
						(e.g., filter down movies/news/products to recommend to user), we expect our
						single CPU-GPU ScratchPipe to be widely applicable as-is over a
						variety of use cases, providing significant reduction in TCO vs. multi-GPUs.  

{\bf ScratchPipe for multi-GPU training.} In
\sect{sect:eval_cost_new}, we demonstrated the advantages of employing single GPU
	ScratchPipe in terms of TCO reduction vs. multi-GPU systems. While this paper
		focuses on the single GPU based ScratchPipe design point, for the
		completeness of our study, we nonetheless discuss the practicality of
		extending our solution for multi-GPU systems below.  A popular approach in
		partitioning and parallelizing RecSys among multi-GPUs is to employ
		table-wise model-level parallelism (i.e., partition distinct set of
				embedding tables across different GPUs) and have each GPU locally go
		through the embedding layer forward/backpropagation as usual, i.e., each
		partitioned embedding table is locally treated as an \emph{independent}
	embedding table from each GPU's perspective~\cite{mudigere2021high}. ScratchPipe is designed to handle GPU cache management in a per embedding table granularity,
	 i.e., RecSys with $N$ embedding tables will have $N$ instance of
		 ScratchPipe's cache manager module instantiated.  As existing
		 model-parallelization for embeddings are carefully designed to
		 enable each GPU to locally handle forward and backpropagation of
		 embedding layers independently with minimum communication, there
		 is no further inter-GPU RAW hazards or any reordering of embedding lookup
		 indices
		 required for correctness and we expect that ScratchPipe can be
		 seamlessly integrated for multi-GPU training.
That being said, while extending ScratchPipe for multi-GPU systems is a viable
design point to explore, it is likely not going to be cost-effective 
in terms of TCO reduction. This is because
parallelizing the DNNs over multi-GPUs only
provides incremental end-to-end performance improvements
(i.e., even with a single GPU, the DNNs are not the
most crucial bottleneck, \fig{fig:eval_latency_breakdown}). 
Consequently, ScratchPipe over multi-GPUs can severely underutilize the
abundant GPU compute throughput, leading to lower training
cost savings.
A detailed,
quantitative evaluation of such is beyond the scope of this paper
and we leave it as future work.

\section {Conclusion}
\label{sect:conclusion}

This paper proposes \proposed, our unique GPU scratchpad based pipelined
\recsys training system.  We first provide a detailed characterization 
 on the locality properties of real-world \recsys datasets, 
 root-causing embedding layer training as a key performance
bottleneck.  We then present several of our unique observations that motivate
our \proposed design, e.g., using the \recsys training dataset to precisely
estimate the past, current, and future embedding table access patterns. Our GPU
scratchpad memory utilizes such property to proactively prefetch
soon-to-be-accessed embeddings into GPU memory, enabling an embedding cache
architecture that always hits, drastically reducing the time spent conducting
embedding layer training for \recsys.  We evaluate our purely software-based
\proposed  over real systems, achieving an average $2.8\times$ (max $4.2\times$) speedup vs.  prior GPU embedding
cache.

\section*{Acknowledegment}
This research is partly supported by 
the National Research Foundation of Korea (NRF)
	grant funded by the Korea government(MSIT) (NRF-2021R1A2C2091753),
the Engineering Research
Center Program through the NRF funded
by the Korean government MSIT under grant NRF-2018R1A5A1059921, and by
Samsung Advanced Institute of Technology (SAIT).
We also appreciate the support from Samsung Electronics Co., Ltd.
Minsoo Rhu is the corresponding author.

\bibliographystyle{ieeetr}
\bibliography{references}

\end{document}